\documentclass[%
 superscriptaddress,
reprint,
 amsmath,amssymb,
 aps,longbibliography
]{revtex4-1}

\usepackage{graphicx}
\usepackage{dcolumn}
\usepackage{bm}
\usepackage{hyperref}
\usepackage[mathlines]{lineno}
\usepackage[bottom]{footmisc}
\usepackage{booktabs}
\makeatletter
\renewcommand*{\@fnsymbol}[1]{\ensuremath{\ifcase#1\or *\or ** \or \ddagger\or
    \mathsection\or \mathparagraph\or \|\or **\or \dagger\dagger
    \or \ddagger\ddagger \else\@ctrerr\fi}}
\makeatother

\begin{document}

\title{Compact dual-band multi-focal diffractive lenses}

\author{Wesley A. Britton}
\thanks{These authors contibute equally to this work}
\affiliation{Division of Materials Science \& Engineering, Boston University, 15 Saint Mary's St., Brookline, Massachusetts 02446, USA}

\author{Yuyao Chen}
\thanks{These authors contibute equally to this work}
\affiliation{Department of Electrical \& Computer Engineering and Photonics Center, Boston University, 8 Saint Mary's Street, Boston, Massachusetts, 02215, USA}

\author{Fabrizio Sgrignuoli}
\affiliation{Department of Electrical \& Computer Engineering and Photonics Center, Boston University, 8 Saint Mary's Street, Boston, Massachusetts, 02215, USA}

\author{Luca Dal Negro}
\email[email:]{dalnegro@bu.edu}
\affiliation{Division of Materials Science \& Engineering, Boston University, 15 Saint Mary's St., Brookline, Massachusetts 02446, USA}
\affiliation{Department of Electrical \& Computer Engineering and Photonics Center, Boston University, 8 Saint Mary's Street, Boston, Massachusetts, 02215, USA}
\affiliation{Department of Physics, Boston University, 590 Commonwealth Avenue, Boston, Massachusetts, 02215, USA}

\begin{abstract}
We design, fabricate, and characterize multifunctional and compact diffractive microlenses with sub-wavelength thickness and the capability to simultaneously focus visible and near-infrared spectral bands at two different focal positions with 24\% and 15\% measured focusing efficiency, respectively. Our technology utilizes high-index and low-loss sputtered hydrogenated amorphous Si, enabling a sub-wavelength thickness of only  235\,nm. Moreover, the proposed flat lens concept is polarization insensitive and can be readily designed to operate across any desired wavelength regime. Imaging under broadband illumination with independent focal planes for two targeted spectral bands is experimentally demonstrated, enabling the encoding of the depth information of a sample into different spectral images. In addition, with a small footprint of only 100\,$\mu$m and a minimum feature size of 400\,nm, the proposed multifunctional and compact diffractive microlenses can be readily integrated with vertical detector arrays to simultaneously concentrate and spectrally select electromagnetic radiation. This provides novel opportunities for spectroscopic and multispectral imaging systems with advanced detector architectures.
\end{abstract}

\maketitle
\section{Introduction}
Spatial and spectral discrimination of electromagnetic radiation are the basic mechanisms that enable spectroscopic and multispectral imaging devices. Multispectral systems are essential tools in medical diagnostics \cite{lu2014medical}, environmental sensing, \cite{kasischke1997use} and defense technologies \cite{shimoni2019hypersectral}. However, on-chip integration of advanced multispectral systems requires significant improvements in miniaturization as well as the demonstration of novel optical functionalities that presently drive the impressive development of metalenses and flat optics technologies  \cite{herzig2014micro,kress2009applied,capasso2018future,genevet2017recent,banerji2019imaging}. The flat optics approach has demonstrated exceptional control and engineering of optical waves with capabilities that are often unattainable using alternative designs or naturally available materials \cite{hu2016metamaterial,cheben2018subwavelength,romagnoli2018graphene,redding2013compact,boschetti2019spectral,kildishev2013planar,koenderink2015nanophotonics,zheludev2012metamaterials}. In particular, flat optics microlenses have been demonstrated using sub-wavelength resonant elements that impart desired phase profiles on a single plane, following the paradigm of metasurfaces \cite{yu2011light,yu2014flat}, or using optimally designed nanostructured masks that focus light beyond the diffraction limit based on optical super-oscillations \cite{zhang2008superlenses,rogers2012super,lalanne2017metalenses}.
Specifically, phase-engineered metalenses and/or numerically optimized super-oscillation lenses achieve desired phase profiles within optically flat devices \cite{zhu2017traditional, kuznetsov2016optically}
for the realization of photonic components with novel functionalities and sub-wavelength thicknesses \cite{arbabi2015subwavelength,khorasaninejad2016polarization,chen2019broadband}. However, most meta-optics designs achieve their desired phase profiles through either geometrical phase modulation, which requires polarized radiation \cite{lin2014dielectric,zheng2015metasurface}, or through engineered resonance behavior, which introduces unavoidable losses that reduce the overall focusing efficiency \cite{zhan2016low,khorasaninejad2016polarization}. Furthermore, the development of meta-optical components with deep sub-wavelength elements for visible applications requires specialized nanofabrication with reduced scalability, especially when non-conventional phase gradients and modulations are required to achieve novel optical functionalities beyond single-point imaging. 

Alternatively, multifunctional diffractive microlenses with engineered electromagnetic phase distributions are well-suited for on-chip photonic integration, have demonstrated excellent scalability \cite{fleming1997blazed} and have efficiencies and imaging performances that often exceed current meta-optics approaches \cite{banerji2019imaging, lalanne2017metalenses}. For example, multilevel diffractive lenses (MDL) can be conveniently produced by photolithography and can reach first-order diffraction efficiencies $>$ 90\% after only few processing iterations \cite{swanson1989binary}. In addition, MDLs with high numerical aperture (NA) have been demonstrated with values as large as 0.9 \cite{mohammad2018broadband} and 1.43 \cite{chao2005immersion} when immersed in water.

However, traditional diffractive microlenses are limited to operation within a single narrow wavelength band. Moreover, the first-order diffraction efficiencies can decrease by as much as 50\% over their fractional bandwidth \cite{swanson1989binary}. Likewise, the intrinsically large focusing dispersion of diffractive microlenses significantly limit their broadband performance. More advanced phase engineering approaches seek to decrease MDL dispersion resulting in achromatic focusing behavior \cite{mohammad2018broadband,davidson1991holographic}. 

In this paper, we design, fabricate, and characterize four-level diffractive microlenses with engineered phase profiles that allow for selective focusing of two well-separated wavelength bands in the visible and near-infrared regime. The proposed multifunctional design provides extended capabilities to diffractive microlenses beyond single-band wavelength applications. Our fabricated compact dual-band multi-focal diffractive lenses have sub-wavelength thicknesses and a small footprint of $100 \mu m$, facilitating integration with compact on-chip elements and microlens arrays \cite{tang2019dual,5401117,lutz2018towards, gunapala2010demonstration,davies1994design,nussbaum1997design}. Moreover, the use of a high-index hydrogenated amorphous Si (a-Si:H) platform is well-suited for monolithic integration with vertical point detectors or detector arrays, enabling compact spectroscopic technologies and advanced multispectral imaging systems.
\section{Design and simulation method}

\begin{figure}[t!]
\centering
\includegraphics[width=\linewidth]{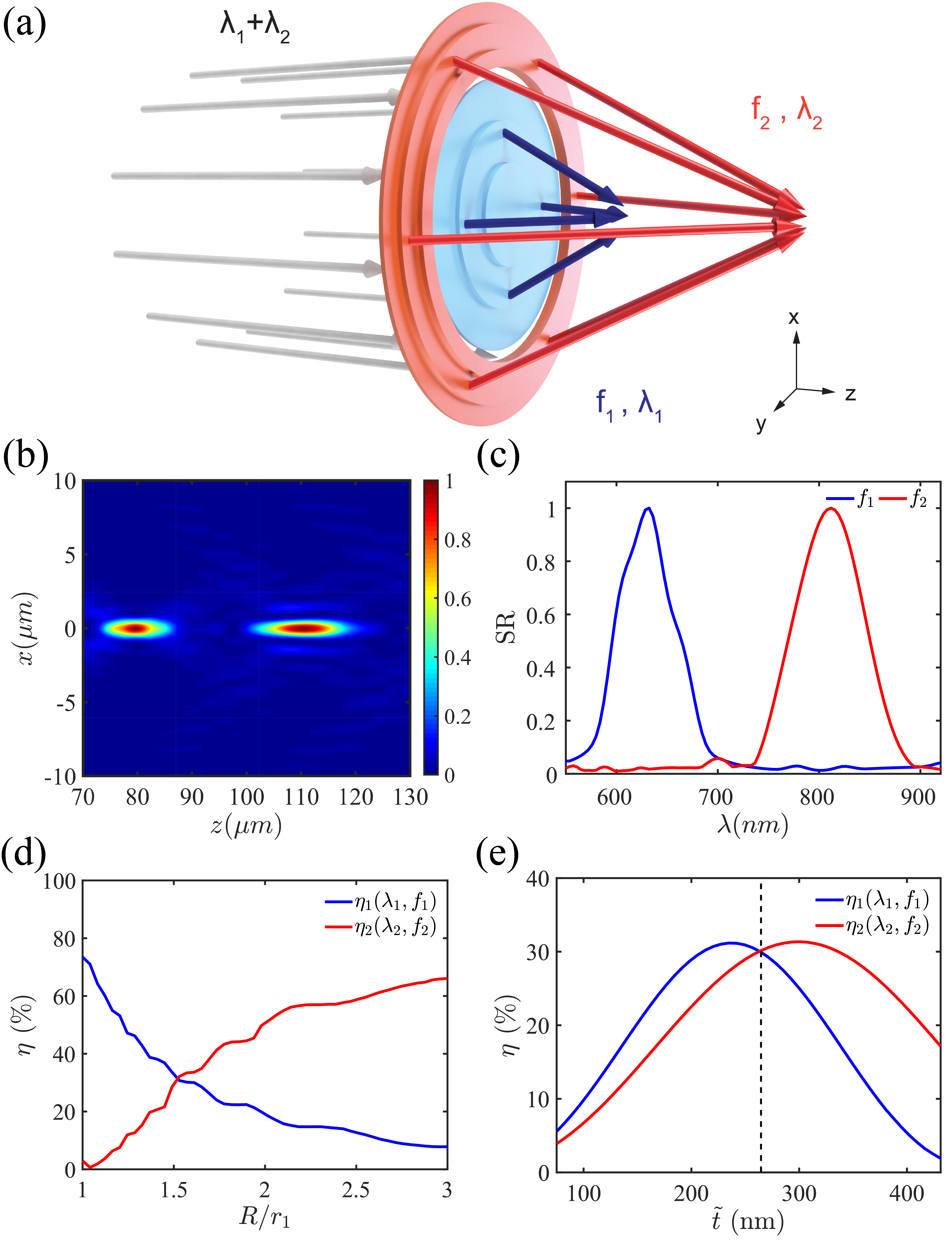}
\caption{(a) Schematic of flat dual-band multi-focal lenses. These devices are characterized by two concentric regions (the blue and the orange one) with tailored spatial distributions that, upon illumination with a polychromatic spectrum, will focus selected wavelength bands centered at $\lambda_1$ and $\lambda_2$ at designed focal points $f_1$ and $f_2$. (b) Simulated intensity profile, normalized with respect to its maximum value, in the $xz$-plane produced by a MDL with $R/r_1$ equal to 1.71. $R$ and $r_1$ are the radius of the inner and outer region, respectively. (c) Normalized spectral response of the dual-band MDL device discussed in panel (b) evaluated at $f_1=80\,\mu$m and $f_2=110\,\mu$m. (d) Computed focusing efficiency $\eta$ for the two designed wavelengths as a function of the geometrical parameter $R/r_1$. (e) Design of the focusing efficiencies $\eta_1$ and $\eta_2$, by fixing $R/r_1$ equals to 1.55, as a function of different effective thicknesses $\tilde{t}$.}
\label{Fig1}
\end{figure}

Figure\,\ref{Fig1}(a) illustrates the basic concept of dual-band multi-focal compact diffractive lens elements. These devices are based on two distinct phase modulations spatially separated into an inner circular region, shown in light blue, and an outer annular region, shown in orange, interlocked on the same device plane. 
Each phase component is responsible for focusing a distinct wavelength band at a selected focal position along the optical $z$-axis. Namely, upon illumination with a polychromatic spectrum this device will focus selected wavelength bands centered around $\lambda_1$ and $\lambda_2$ at pre-defined distances $f_1$ and $f_2$ along the $z$-axis. This operation mechanism differs from traditional multi-focal lenses that generate multiple longitudinal focal spots for the same wavelength \cite{chen2015longitudinal,williams2017plasmonic}. Rather, our device design enables the simultaneous focusing of different spectral bands either at separate focal points or at the same one. 

\begin{figure*}[t]
\centering
\includegraphics{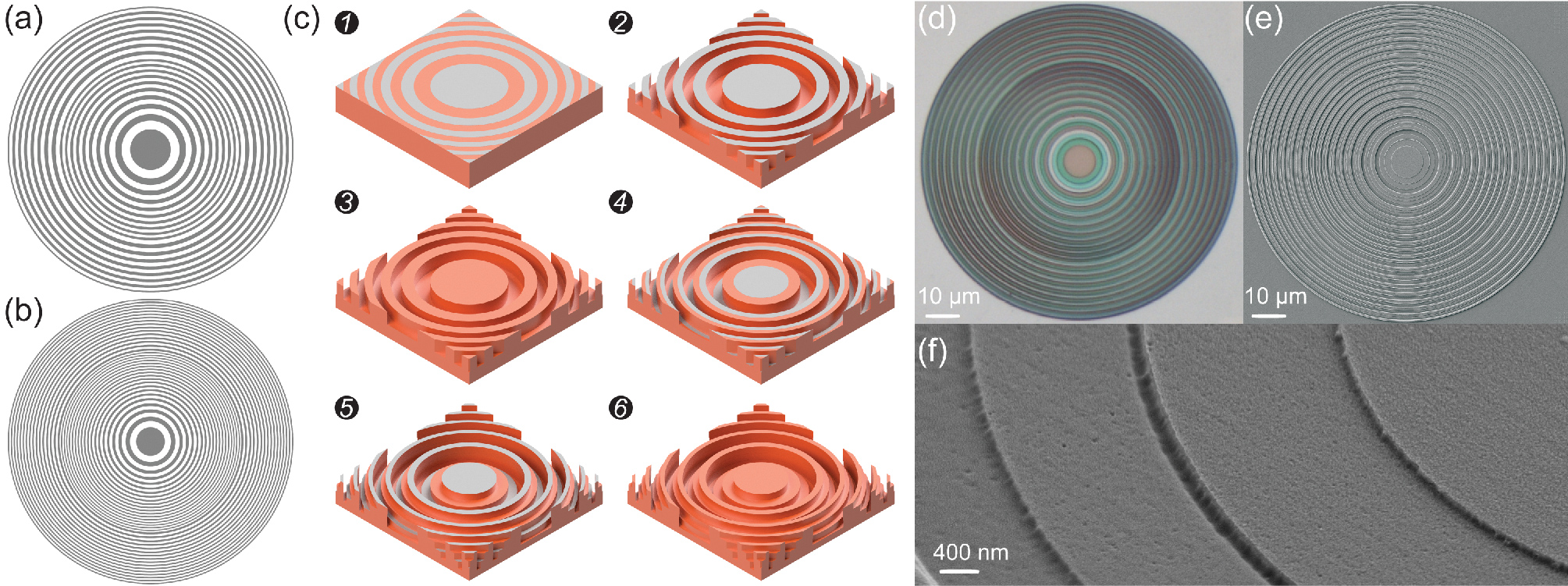}
\caption{First (a) and second (b) lithographic masks used to fabricate the dual-band MDL microlens. (c) Fabrication process flow based on top-down lithography methods. (d) Top-view of the optical reflection image of the fabricated device upon bright field illumination. Panels (e) and (f) show, respectively, the full top-view and a higher magnification at 45$^\circ$ of the SEM images of the fabricated MDL microlens.}
\label{fig:2}
\end{figure*}

We study the wave diffraction problem of the proposed dual-band multi-focal MDL using the rigorous Rayleigh-Sommerfeld (RS) first integral formulation. This approach is valid beyond the paraxial approximation \cite{goodman2005introduction}. We consider devices located on the $xy$ plane illuminated by a normally incident plane wave, which results in a diffracted field that propagates along the $+z$ direction. The field amplitude is computed according to RS with the following expressions:
\begin{eqnarray}\label{eqn:RS 1}
    U_{2}\left(x^\prime,y^\prime;z,k\right)=U_{1}\left(x,y\right) * h(x,y;z,k)\\
    \label{eqn:RS 2}
    h(x,y;z,k)=\frac{1}{2\pi}\frac{z}{\rho}  \left(\frac{1}{\rho}-jk\right)\frac{e^{\left(jk\rho\right)}}{\rho}.
\end{eqnarray}
Here $*$ denotes the convolution operation, while $U_{1}$ and $U_{2}$ are the complex transverse field distributions of the object and of the image, respectively. Furthermore, $k$ is the incident wavenumber, and $\rho=\sqrt{x^2+y^2+z^2}$, where $z$ is the distance between the device and image plane. We consider $U_{1}\left(x,y\right)=e^{j\Phi(r)}$, where $\Phi(r)$ is the phase profile on the device that is explicitly represented as: 
\begin{equation}
\Phi(r) =
\begin{cases} 
      \Bigg\{-2\pi\left[\left(\sqrt{r^2+f_1^2}-f_1\right)\right]/\lambda_1\Bigg\}_{2\pi}& 0<r<r_1\\
      \Bigg\{-2\pi\left[\left(\sqrt{r^2+f_2^2}-f_2\right)\right]/\lambda_2\Bigg\}_{2\pi} & r_1<r<R.\\
   \end{cases}
   \label{eqn:phase profile1}
\end{equation}
where $r_1$ and $R$ indicate the maximum radial extensions of the first and of the second phase components, respectively. We utilize the phase profiles of separate Fresnel lenses because they guarantee large focusing efficiency around each of the operating wavelengths $\lambda_1,\lambda_2$ of the two targeted spectral bands and at the focal lengths $f_1,f_2$. The $2\pi$ subscripts in eq.\,(\ref{eqn:phase profile1}) indicate the reduction by modulo $2\pi$.  This phase design retains rotational symmetry along the optical axis, which preserves polarization independence. Likewise, it ensures a rotationally symmetric point spread function (PSF) for improved focusing and imaging quality. In the present study, we have discretized the phase profile $\Phi(r)$ into four levels, which significantly improves ease of fabrication by requiring only two different lithographic steps.

As a proof of principle, we have selected visible and near-infrared (NIR) spectral regions centered at $\lambda_1 = 632.8$\,nm and $\lambda_2 = 808$\,nm. In addition,  we have chosen the small focal distances $f_1 = 80\,\mu$m and $f_2 = 110\,\mu$m, which are compatible with on-chip multi-layer detector technologies \cite{tang2019dual,5401117}. We remark that by changing the radial extension of the first and second phase components, we can control both their relative intensity and focusing efficiency. Indeed, by choosing $R/r_1=1.71$ , we can design a device with equal focusing intensities at two well-separated focal lengths. This is shown in panel (b) where we report the intensity distribution profile of the two focal spots on the $xz$-plane. Furthermore, its spectral response (SR), defined as the maximum field intensity at the two designed focusing planes ($f_1=80\,\mu$m and $f_2=110\,\mu$m) as a function of wavelengths, is shown in Fig.\,\ref{Fig1} (c). Our results demonstrate that the two wavelength bands are not only well-separated spatially, but also spectrally. This feature is important in multispectral imaging applications \cite{tang2019dual,5401117} and for the integration with optoelectronic devices \cite{arbabi2015subwavelength}. Moreover, we can also independently control the focusing efficiencies $\eta_1$ and $\eta_2$ of the inner and outer phase components through the geometrical ratio $R/r_1$. This is demonstrated in Fig.\,\ref{Fig1}\,(d). In our work, we define the focusing efficiency $\eta(\lambda,z)$ as the ratio of the image plane intensity $I'$ within a spot size of three times the full-width-at-half-maximum (FWHM) to the incident intensity profile $I$ integrated over the entire device area \cite{banerji2019imaging,arbabi2015subwavelength}:

\begin{equation}
\eta(\lambda,z) = \frac{\int_{0}^{3FWHM}dr^\prime\int_{0}^{2\pi}d\theta^\prime I^\prime(\lambda,z,r^\prime,\theta^\prime)}{\int_{0}^{R}dr\int_{0}^{2\pi}d\theta I(\lambda,z=0,r,\theta)}.
   \label{eqn:focusing efficiency}
\end{equation}

From Fig.\,\ref{Fig1}\,(d), evaluated considering $I'(\lambda,z,r',\theta')=|U_2|^2$, we find that a value of $R/r_1=1.55$ equalizes the two efficiencies such that $\eta_1(\lambda_1,f_1)=\eta_2(\lambda_2,f_2)= 31\%$. Note that, in addition to retaining the previously defined parameter values, we have also set $R=50\,\mu$m (see \href{link}{Supplement 1} for more details regarding the dependance of the device performance on $R$). We selected $r_1\approx31\,\mu$m and $R=50\,\mu$m to fabricate and experimentally characterize the designed compact microlens.

We recall that MDLs exploit variations in the thickness of materials to impart desired phase delays on the incident radiation. In order to satisfy the designed phase condition of our dual-band device, the material thickness of the inner region $t_1$ and outer annular region $t_2$ have to be equal to $t_{k}= 3\lambda_{k}/4(n-1)$, where $n$ is refractive index and $k=1,2$. Therefore, four lithography process steps are required to fabricate the resulting device. However, to ease this fabrication process, we can identify a central wavelength $\tilde{\lambda}$ to determine an effective total device thickness $\tilde{t}$ =$3\tilde{\lambda}/4(n-1)$ across the entire device area, reducing the required lithographic process steps to only two with a small reduction in the overall efficiency. This is reported in Fig.\,\ref{Fig1}\,(e) where we display the focusing efficiency of the inner ($\eta_1$ in blue line) and outer ($\eta_2$ in red line) phase components as a function of $\tilde{t}$. Specifically, an effective device thickness equal to 266\,nm, indicated by the black dashed line in Fig.\,\ref{Fig1}\,(e), equalizes $\eta_1$ and $\eta_2$ to 30\%, resulting in a reduction of only 1$\%$ with respect to the results of panel (d).
\section{Growth and fabrication method}
\begin{figure*}[t!]
\centering
\includegraphics{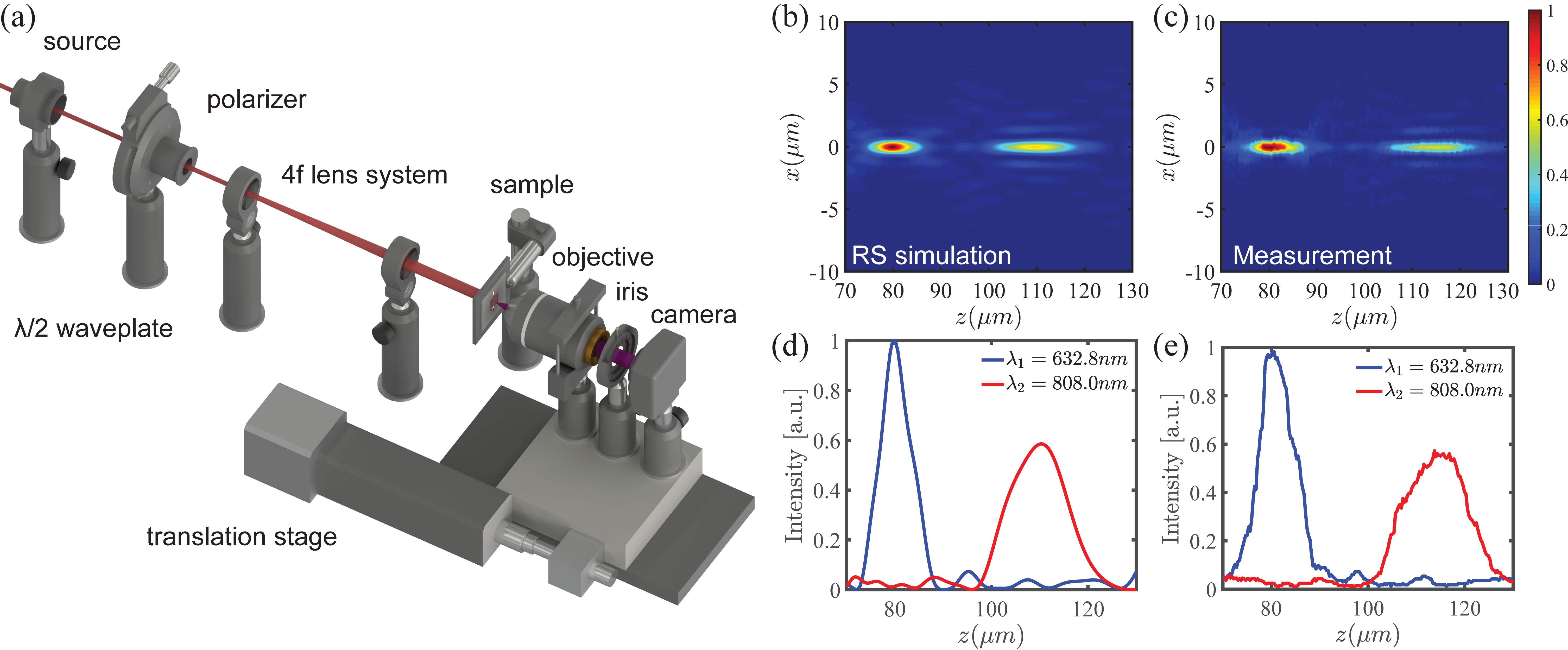}
\caption{(a) Experimental optical setup used for measuring the focusing properties of the fabricated multilevel diffractive lenses. Simulated (b) and (c) measured intensity profiles along the focusing $xz$-plane normalized with respect to its maximum. Panels (d) and (c) show the intensity cut-profile of the normalized maps (b) and (c) as a function of the optical axis $z$, respectively.}
  \label{fig:3}
\end{figure*}
Reactively sputtered hydrogenated amorphous Si (a-Si:H) enables the compact realization of our design. This material has excellent transparency and a high refractive index ($n\approx3.0$) at visible and NIR wavelengths. Likewise, physical deposition by sputtering is a low cost and scalable process. Markedly, well-hydrogenated and alloyed low-loss a-Si:H is transparent until a sharp increase in the optical absorption edge between 2.1 $eV$ and 2.2 $eV$, representing an optical bandgap \cite{moustakas1979sputtered}. However, sputtered a-Si:H dispersion and optical properties are highly dependent on deposition conditions. The a-Si:H thin films in this study were grown reactively on fused silica, SiO$_2$, substrates using a Denton Discovery D8 magnetron sputtering system. In preparation before deposition, substrates were solvent washed, sonicated, and plasma ashed in an O$_2$ atmosphere (M4L, PVA TePla America). The a-Si:H films were deposited using the following conditions: sputtering base pressure was kept below $3\times10^{-7}$ Torr, deposition pressure was kept at $10$ mTorr with a Ar:H$_2$ gas flow ratio of 4:1 Sccm, substrate was heated to $300^\circ$C, and RF deposition power on the $3''$ intrinsic Si target was set to 200W. The materials' optical dispersion was characterized using ellipsometry (V-VASE, J.A. Woollam) on 50 nm films. Our deposition process was optimized so that the final films used for fabrication have an average index of $n=3.0$ and absorption coefficient of $\alpha<5\times10^3$cm$^{-1}$ across the entire operational range. Detailed film characterization is provided in the \href{link}{Supplement 1}. 

As discussed in the previous section, we identify an effective total device thickness $\tilde{t}$. As a consequence, our fabrication method uses only two lithography process steps to realize a four-level discretized device. Figure\,\ref{fig:2}\,(a-b) show the first and second etch mask designs, respectively. The first etch mask is created by combining the $3\pi/2$ and $\pi$ level phase areas, and the second etch mask is the combination of the $3\pi/2$ and $\pi/2$ level areas. The general fabrication process flow is shown in Fig.\,\ref{fig:2}\,(c). Steps 1 through 3 and 4 through 6 are associated with the first and second process iterations, respectively. First, steps 1 and 4 are the lithographic patterning of the first and second etch mask, respectively. The designed etch masks are transferred to metal hard masks with lithography and electron beam evaporation using a positive resist and a lift-off process. Next, steps 2 and 5 are deep anisotropic etches of thicknesses with target values $\tilde{\lambda}/2(n-1)=177$\, nm and $ \tilde{\lambda}/4(n-1)=89$\,nm, respectively.  Finally, steps 3 and 6 are wet etch removals of the residual hard masks. 

Our specific fabrication process utilizes electron beam lithography (EBL), a Cr hardmask, and a reactive ion etching (RIE) anisotropic dry etch for both processing iterations. In addition, we deposit Au alignment marks with Ti adhesion layers before device fabrication. For the lithography process, MicroChem PMMA A3 positive resist was spun at 3000 rpm and baked in an oven at 170$^\circ C$ for 20 min. After both resist baking steps, a thin conducting layer of Au was deposited using a Cressington 108 Manual Sputter Coater. Details on the exposure and etching processes of our device are reported in the \href{link}{Supplement 1}.

Figure\,\ref{fig:2}\,(d-e) show optical and SEM images of the top view of the fabricated device. Each of the four phase layers in Fig.\,\ref{fig:2} (d) are distinguishable as different colors because of color enhancement by optical interference effects. The phase jump at $r = r_1$ is clear in both of these images. A higher magnification $45^\circ$ angled SEM image of the fabricated device is shown in Fig.\ref{fig:2}\,(f). This image shows all four levels of the fabricated device. However, due to the etch rate reduction associated to the RIE lag, the total device thickness of the fabricated device results to be 235\,nm \cite{jansen1997bsm} (see also \href{link}{Supplement 1} for more details).

\section{Optical characterization method}
The performances of the fabricated devices are experimentally characterized using the customized visible and near-infrared (NIR) imaging system with $z$-translation capabilities, as illustrated in Fig.\ref{fig:3}\,(a). It utilizes a tunable Ti:sapphire (Mai Tai, Spectra-Physicsop) quasi-CW laser source ($fs$ pulse width, but with MHz repetition rate) that allows wavelength selective measurements over a broad visible and NIR spectral band (690\,nm -- 1000\,nm), and a HeNe laser (Hughes 3225H-PC) for characterization at 632.8\,nm. We incorporate a $\lambda/2$ waveplate and a linear polarizer to attenuate both sources. In addition, a $4f$ lens system is used to expand the beam, which approximates a plane wave source illumination and reduces any contribution from beam convergence or divergence. An imaging system consisting of a 0.9 numerical aperture (NA) objective (Olympus MPLFLN100xBDP), an iris (to remove stray reflections), and a CMOS camera (Thorlabs DCC1645C, with short-pass filter removed) is translated along the $z$-axis in unison by a motorized translation stage (Thorlabs LNR50S). This allows the characterization of the image plane intensity profile as a function of distance along the $z$-axis.

The $x$- and $y$-axis scaling was calibrated using an Au alignment mark with known dimensions. The reference data was measured by a transverse translation away from the device and imaging the bare substrate and residual a-Si:H film. Reference and signal data were taken at different exposure times to compensate for the increase in intensity due to focusing. The dark background average was subtracted from both data sets to compensate for the different exposure times. 
\section{Results and discussion}\label{Results and discussion}

Our dual-band multi-focal compact microlens was initially characterized at its operation wavelengths, $\lambda_1$= 632.8\,nm and $\lambda_2$= 808\,nm. Figure\,\ref{fig:3}\,(b) shows the expected electric field intensity distribution in the $xz$-plane, normalized with respect to its maximum. This plot shows that the two targeted wavelengths are both well focused at their designed focal distances $f_1=80\,\mu$m (thus a NA of 0.4) and $f_2=110\,\mu$m (thus a NA of 0.47), respectively. Figure\,\ref{fig:3}\,(c) displays the measured intensity distribution in the $xz$-plane, also normalized with respect to its maximum. This analysis was performed by combining a series of images taken parallel to the device plane when scanning the $z$-axis with a resolution of 0.3 $\mu$m. The measurements are in excellent agreement with the simulations demonstrating the validity of the proposed approach. In order to compare the position of the maximum value of the focusing intensity along the optical axis as well as the focal depth obtained in the experiments with respect to simulation, we report their intensity cut-profile along the $z$-axis in Figs.\,\ref{fig:3}\,(d) and (e), respectively. The measured FWHM of the focal depths of the two phase components, shown in Fig.\,\ref{fig:3}\,(e), are found to be  equal to $14\,\lambda_1$ and $20\,\lambda_2$, respectively. These values are close to the predicted ones calculated to be $12\,\lambda_1$ and $18\,\lambda_2$, respectively (see Fig.\,\ref{fig:3}\,(d)). On the other hand, we measured an efficiency of $\eta_1(\lambda_1,f_1)= 24\%$ and $\eta_2(\lambda_2,f_2)= 15\%$ rather than $\eta_1(\lambda_1,f_1)=\eta_2(\lambda_2,f_2)=30\%$. This difference in efficiency is due to the small discrepancy in the thickness of the fabricated device with respect to the designed $\tilde{t}$, which is caused by the etch rate variations at small feature sizes \cite{jansen1997bsm}. Specifically, the total thickness of the fabricated device deviates by approximately 30\,nm  from the designed value of 266\,nm that  equalizes $\eta_1$ and $\eta_2$. Given the thickness of the fabricated device we can estimate, based on the RS calculations reported in Fig.\,\ref{Fig1}\,(e), that the two focusing efficiencies will be different, \emph{i.e.} $\eta_1(\lambda_1,f_1)=31\%$ and $\eta_2(\lambda_2,f_2)=26\%$. However, RS simulations, while enabling the analysis of large-scale devices, are based on scalar diffraction theory. Therefore, these simulations do not account for vector coupling effects that become important when the incident radiation interacts with the features of the device that are comparable in size or smaller than the wavelength, resulting in a wavelength-dependent efficiency drop. Often referred to as ''light shadowing effects'', these vector diffraction effects  include reflection losses and the resonant behavior observed in devices with high numerical apertures  \cite{lalanne1999design,golub2007analytic}. See Supplement 1 for more details regarding full-vector simulations using the Finite Element Method (FEM). 

\begin{figure}[t]
\centering
\includegraphics[width=\linewidth]{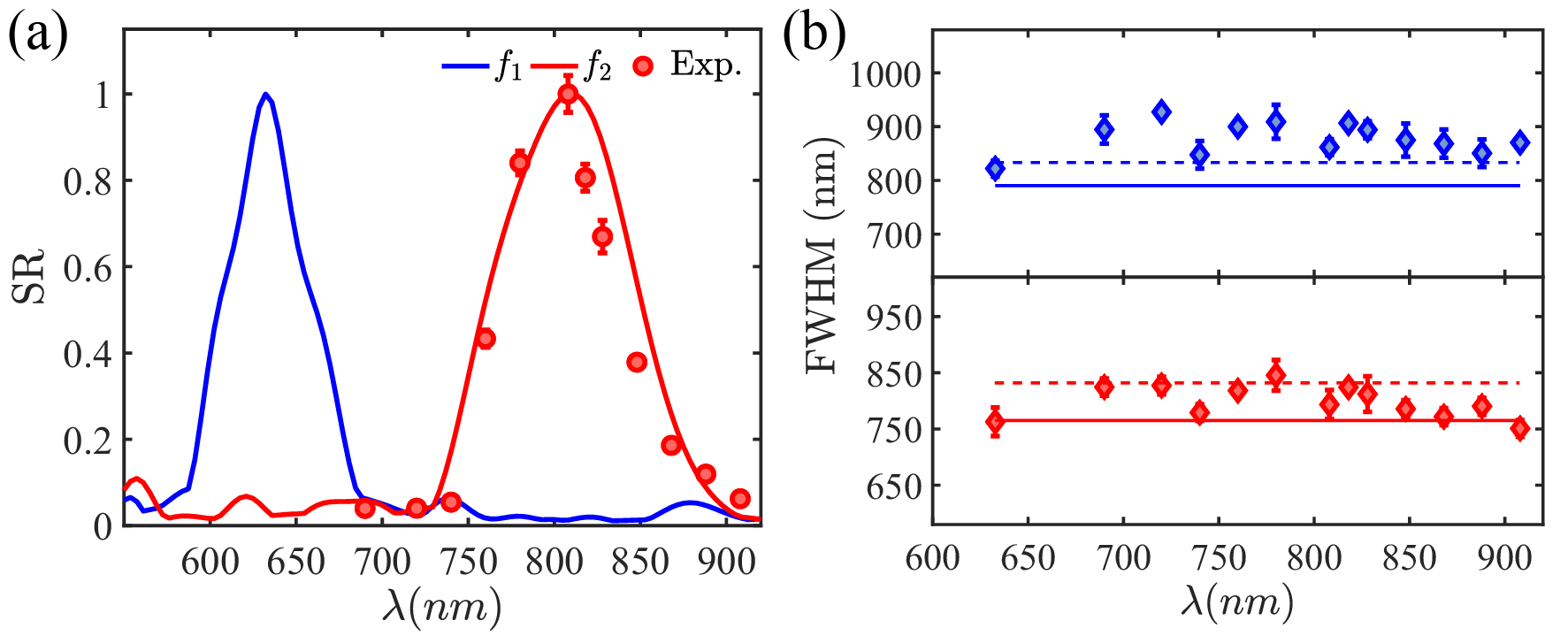}
\caption{Panel (a) shows the simulated spectral response of the MDL device evaluated at the focal plane $f_2$=110$\,\mu$m (continuous red-line), normalized with respect to its maximum value. The measured SR is also reported with red-circle markers. Panel (b) compares the measured (diamond-markers) and the simulated (dashed-lines) FWHM of the point spread functions relative to the two phase components (blue for inner component and red for the  outer component), respectively, with respect to the analytical diffraction-limited PSF (continuous-lines). The error bars are evaluated by using the results obtained at the planes one scanning step before and after the focal plane.}
  \label{fig:4}
\end{figure}

\begin{figure*}[t]
\centering
\includegraphics{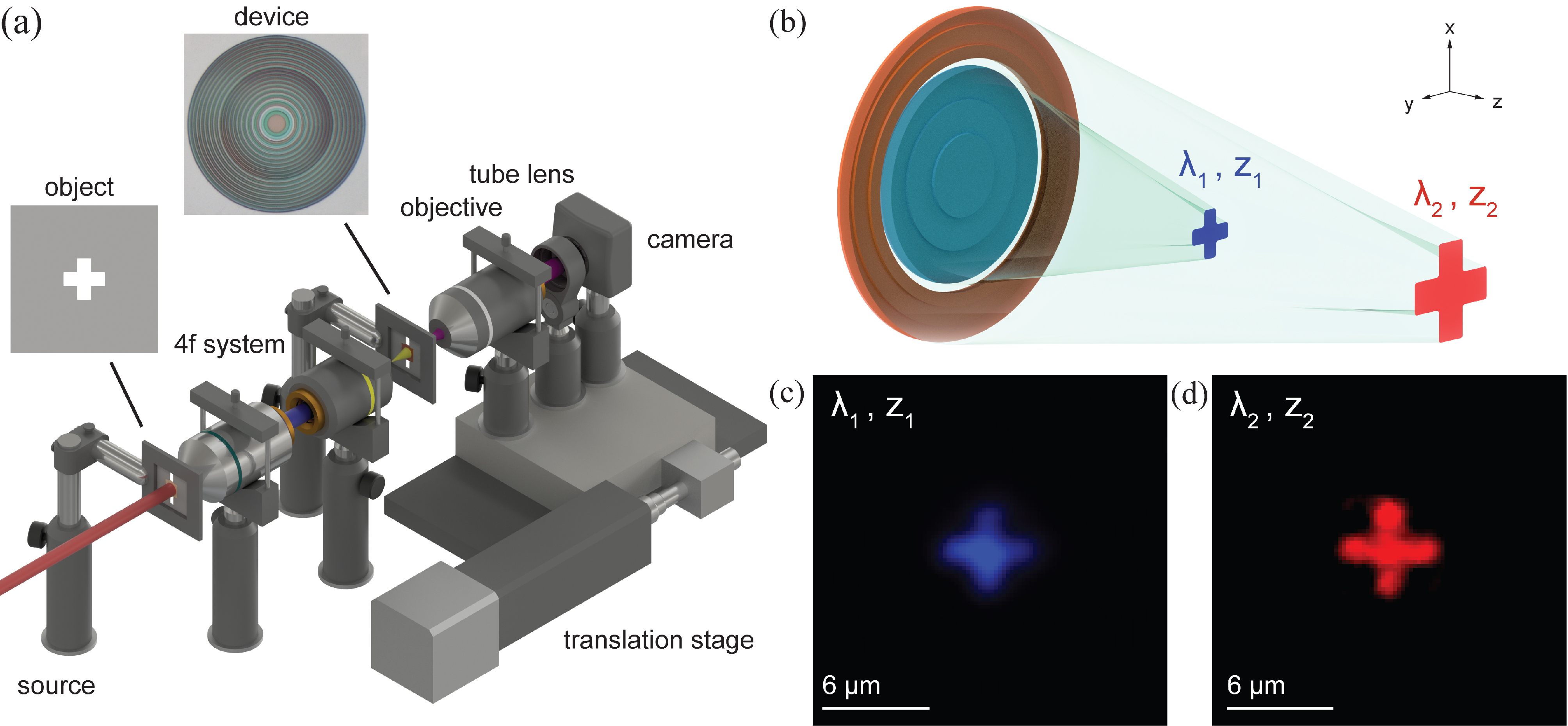}
\caption{(a) Experimental optical setup used for measuring the imaging properties of the fabricated multilevel diffractive lenses. (b) Schematic of the imaging behavior of the dual-band multi-focal lens. An object with a polychormatic spectrum will focus the image corresponding to the selected wavelengths bands centered at $\lambda_1$ and $\lambda_2$ at separated image planes at $z_1$ and $z_2$. (c-d) images of the object formed at $z_1$ and $z_2$, respectively.}
  \label{fig:5}
\end{figure*}

In order to further characterize the dual-band mechanism, we display in Fig.\,\ref{fig:4}\,(a) both the normalized measured and simulated spectral response.The experimental SR at $f_1$ is not shown since most of its spectral bandwidth lies outside our laser source wavelength range. On the other hand, we have verified that the spectral bandwidth of the second focus at $f_2$ is consistent between experiment and measurement. Indeed, the simulated and measured FWHM of these spectral responses are equal to 120\,nm and 100\,nm, respectively.

Moreover, we characterized the FWHM of the two focal spots in Fig.\,\ref{fig:4} (b). This parameter quantifies the ability of our device to concentrate electromagnetic radiation. Specifically, the diffraction-limit of a annular focusing system is described by \cite{rivolta1986airy}:
\begin{equation}\label{eqn:efficiency}
I'(\beta(r)) = \frac{I_\circ}{(1-\epsilon^2)^2}\left(\frac{2J_1(\beta(r))}{\beta(r)}-\frac{2\epsilon J_1(\epsilon \beta(r))}{\beta(r)} \right)^2\\
\end{equation}
where $\epsilon= r_1/R$, $\beta(r)= 4{\pi}Rr/{\lambda_2}\sqrt{r^2+f_{2}^2}$, and $J_1$ is the first order Bessel function of the first kind. This function results in an intensity distribution similar to an Airy pattern but with slightly increased intensity in the side lobes. Indeed, eq.\,\ref{eqn:efficiency} reproduces a standard Airy pattern when  $\epsilon=0$ and substituting $R$, $f_2$, and $\lambda_2$ with $r_1$, $f_1$, and $\lambda_1$ in $\beta(r)$. By solving eq.(\ref{eqn:efficiency}) at different focal planes for  different incident wavelengths as a function of $r$ when $I'(\beta(r))$=0.5, we obtain the analytical FWHM spectrum of the point spread functions of the inner and outer phase component. We compare these analytical results in Fig.\,\ref{fig:4}\,(b) with respect to the simulated (dashed lines) and the measured (markers) FWHM as a function of $\lambda$. The experimental results were found to be in good agreement with the simulated values with only a slightly larger FWHM than theory in both cases. These data demonstrate that dual-band multi-focal diffractive lenses produce almost diffraction-limited focal spots at different operation wavelengths.

\section{Imaging behavior}
There is a growing interest in the engineering of flat optical components with the ability to form well-separated images at different positions along both the longitudinal  \cite{chen2015longitudinal,williams2017plasmonic} and vertical directions \cite{khorasaninejad2016multispectral}. In these devices the positions of the focal points are controlled with the helicity or with the polarization of the incident light. In this section, we demonstrate that the developed multi-focal diffractive lenses provide  wavelength-controlled multi-focal imaging using unpolarized light illumination. This results in the formation of two independent images at two different positions depending on the wavelength, as demonstrated in Fig.\,\ref{fig:5}. In Fig.\,\ref{fig:5}(a) we display a schematic picture of the optical setup used to characterize the dual-band imaging performance of our device. Here, we image our object positioned before the compact dual-band multi-focal MDL using a 4f imaging system consisting of two objectives. The imaging behavior of the MDL as a function of position along the $z$-axis can then be characterized by an imaging setup consisting of a high NA objective, tube lens, and camera all mounted on a motorized translation stage. In our experiment we image a 60\,$\mu$m long cross-shaped object defined in a 300\,$\mu$m by 300\,$\mu$m wide square region obtained using a lift-off process with a 50\,nm thick Cr mask on SiO$_2$. We illuminate our object at $\lambda_1$ and $\lambda_2$ using a halogen lamp filtered by a monochromator (Cornerstone 260) and collimated. The operation principle of the fabricated dual-band multi-focal device is illustrated in Fig.\,\ref{fig:5}\,(b). Considering the geometry of our experimental setup, the images will be formed at the plane positions $z_1$ and $z_2$, as predicted by using the Gaussian lens equation \cite{goodman2005introduction}:
\begin{equation}\label{eqn:imagePlane}
\frac{1}{z_{o}}+\frac{1}{z_{m}}=\frac{1}{f_m}\quad(m=1,2)
\end{equation}
where $z_o$ equal to 350$\mu$m is the object distance. As predicted by eq.\,(\ref{eqn:imagePlane}), Fig.\,\ref{fig:5}(c-d) show two clear images of our object at $z_1$=102\,$\mu$m  and $z_2$=155\,$\mu$m when illuminating with $\lambda_1$ and $\lambda_2$, respectively. These results demonstrate that two images are formed for the two selected narrow bands at designed locations along the longitudinal direction.


\section{Conclusion}
In conclusion, we have designed, fabricated, and characterized compact visible and NIR dual-band multi-focal diffractive microlenses. These devices are based on high-index materials compatible with very large scale integration fabrication methodologies resulting in a considerable reduction in cost, complexity, and size compared to alternative implementations based on the meta-lens approach. The proposed multifunctional microlens design shows the capability to focus two different incident wavelengths at two different focal positions along the optical axis. Moreover, the same design concept demonstrated here can be applied to simultaneously focus predefined spectral bands at the same focal spot. This allows integration atop advanced on-chip detector architectures, enabling ultra-compact multispectral devices for imaging and spectroscopic on-chip applications \cite{tang2019dual}. For instance, our devices could be used as multispectral microlens arrays for differential multiphoton microscopy \cite{hoover2011remote,bewersdorf1998multifocal,beaulieu2020simultaneous}. Finally, we have shown that dual-band multi-focal microlenses can image, under unpolarized illumination, an object at two designed focal positions depending on the incident wavelength. This enables the encoding of the depth information of a sample into different spectral images, \emph{i.e.} multispectral confocality \cite{tearney2002confocal}, which can be exploited for 3D image reconstruction and in future tomographic applications.


\section*{Funding Information}
Army Research Laboratory (ARL) (Cooperative Agreement Number W911NF-12-2-0023); National Science Foundation (NSF) (DMR 1709704).

\section*{Acknowledgements}
L.D.N. would like to thank Prof. Enrico Bellotti for insightful discussions on the presented technology.

\section*{Disclosures}
The authors declare no conflicts of interest.

\bigskip \noindent See \href{link}{Supplement 1} for supporting content.

\section{Supplementary Information}

This will usually read something like: ``Experimental procedures and
characterization data for all new compounds. The class will
automatically add a sentence pointing to the information on-line:


\begin{thebibliography}{53}%
\makeatletter
\providecommand \@ifxundefined [1]{%
 \@ifx{#1\undefined}
}%
\providecommand \@ifnum [1]{%
 \ifnum #1\expandafter \@firstoftwo
 \else \expandafter \@secondoftwo
 \fi
}%
\providecommand \@ifx [1]{%
 \ifx #1\expandafter \@firstoftwo
 \else \expandafter \@secondoftwo
 \fi
}%
\providecommand \natexlab [1]{#1}%
\providecommand \enquote  [1]{``#1''}%
\providecommand \bibnamefont  [1]{#1}%
\providecommand \bibfnamefont [1]{#1}%
\providecommand \citenamefont [1]{#1}%
\providecommand \href@noop [0]{\@secondoftwo}%
\providecommand \href [0]{\begingroup \@sanitize@url \@href}%
\providecommand \@href[1]{\@@startlink{#1}\@@href}%
\providecommand \@@href[1]{\endgroup#1\@@endlink}%
\providecommand \@sanitize@url [0]{\catcode `\\12\catcode `\$12\catcode
  `\&12\catcode `\#12\catcode `\^12\catcode `\_12\catcode `\%12\relax}%
\providecommand \@@startlink[1]{}%
\providecommand \@@endlink[0]{}%
\providecommand \url  [0]{\begingroup\@sanitize@url \@url }%
\providecommand \@url [1]{\endgroup\@href {#1}{\urlprefix }}%
\providecommand \urlprefix  [0]{URL }%
\providecommand \Eprint [0]{\href }%
\providecommand \doibase [0]{http://dx.doi.org/}%
\providecommand \selectlanguage [0]{\@gobble}%
\providecommand \bibinfo  [0]{\@secondoftwo}%
\providecommand \bibfield  [0]{\@secondoftwo}%
\providecommand \translation [1]{[#1]}%
\providecommand \BibitemOpen [0]{}%
\providecommand \bibitemStop [0]{}%
\providecommand \bibitemNoStop [0]{.\EOS\space}%
\providecommand \EOS [0]{\spacefactor3000\relax}%
\providecommand \BibitemShut  [1]{\csname bibitem#1\endcsname}%
\let\auto@bib@innerbib\@empty
\bibitem [{\citenamefont {Lu}\ and\ \citenamefont {Fei}(2014)}]{lu2014medical}%
  \BibitemOpen
  \bibfield  {author} {\bibinfo {author} {\bibfnamefont {Guolan}\ \bibnamefont
  {Lu}}\ and\ \bibinfo {author} {\bibfnamefont {Baowei}\ \bibnamefont {Fei}},\
  }\bibfield  {title} {\enquote {\bibinfo {title} {Medical hyperspectral
  imaging: a review},}\ }\href@noop {} {\bibfield  {journal} {\bibinfo
  {journal} {Journal of biomedical optics}\ }\textbf {\bibinfo {volume} {19}},\
  \bibinfo {pages} {010901} (\bibinfo {year} {2014})}\BibitemShut {NoStop}%
\bibitem [{\citenamefont {Kasischke}\ \emph {et~al.}(1997)\citenamefont
  {Kasischke}, \citenamefont {Melack},\ and\ \citenamefont
  {Dobson}}]{kasischke1997use}%
  \BibitemOpen
  \bibfield  {author} {\bibinfo {author} {\bibfnamefont {Eric~S}\ \bibnamefont
  {Kasischke}}, \bibinfo {author} {\bibfnamefont {John~M}\ \bibnamefont
  {Melack}}, \ and\ \bibinfo {author} {\bibfnamefont {M~Craig}\ \bibnamefont
  {Dobson}},\ }\bibfield  {title} {\enquote {\bibinfo {title} {The use of
  imaging radars for ecological applications review},}\ }\href@noop {}
  {\bibfield  {journal} {\bibinfo  {journal} {Remote sensing of environment}\
  }\textbf {\bibinfo {volume} {59}},\ \bibinfo {pages} {141--156} (\bibinfo
  {year} {1997})}\BibitemShut {NoStop}%
\bibitem [{\citenamefont {Shimoni}\ \emph {et~al.}(2019)\citenamefont
  {Shimoni}, \citenamefont {Haelterman},\ and\ \citenamefont
  {Perneel}}]{shimoni2019hypersectral}%
  \BibitemOpen
  \bibfield  {author} {\bibinfo {author} {\bibfnamefont {Michal}\ \bibnamefont
  {Shimoni}}, \bibinfo {author} {\bibfnamefont {Rob}\ \bibnamefont
  {Haelterman}}, \ and\ \bibinfo {author} {\bibfnamefont {Christiaan}\
  \bibnamefont {Perneel}},\ }\bibfield  {title} {\enquote {\bibinfo {title}
  {Hypersectral imaging for military and security applications: Combining
  myriad processing and sensing techniques},}\ }\href@noop {} {\bibfield
  {journal} {\bibinfo  {journal} {IEEE Geoscience and Remote Sensing Magazine}\
  }\textbf {\bibinfo {volume} {7}},\ \bibinfo {pages} {101--117} (\bibinfo
  {year} {2019})}\BibitemShut {NoStop}%
\bibitem [{\citenamefont {Herzig}(2014)}]{herzig2014micro}%
  \BibitemOpen
  \bibfield  {author} {\bibinfo {author} {\bibfnamefont {Hans~Peter}\
  \bibnamefont {Herzig}},\ }\href@noop {} {\emph {\bibinfo {title}
  {Micro-optics: elements, systems and applications}}}\ (\bibinfo  {publisher}
  {CRC Press},\ \bibinfo {year} {2014})\BibitemShut {NoStop}%
\bibitem [{\citenamefont {Kress}\ and\ \citenamefont
  {Meyrueis}(2009)}]{kress2009applied}%
  \BibitemOpen
  \bibfield  {author} {\bibinfo {author} {\bibfnamefont {Bernard~C}\
  \bibnamefont {Kress}}\ and\ \bibinfo {author} {\bibfnamefont {Patrick}\
  \bibnamefont {Meyrueis}},\ }\href@noop {} {\emph {\bibinfo {title} {Applied
  Digital Optics: from micro-optics to nanophotonics}}}\ (\bibinfo  {publisher}
  {John Wiley \& Sons},\ \bibinfo {year} {2009})\BibitemShut {NoStop}%
\bibitem [{\citenamefont {Capasso}(2018)}]{capasso2018future}%
  \BibitemOpen
  \bibfield  {author} {\bibinfo {author} {\bibfnamefont {Federico}\
  \bibnamefont {Capasso}},\ }\bibfield  {title} {\enquote {\bibinfo {title}
  {The future and promise of flat optics: a personal perspective},}\
  }\href@noop {} {\bibfield  {journal} {\bibinfo  {journal} {Nanophotonics}\
  }\textbf {\bibinfo {volume} {7}},\ \bibinfo {pages} {953--957} (\bibinfo
  {year} {2018})}\BibitemShut {NoStop}%
\bibitem [{\citenamefont {Genevet}\ \emph {et~al.}(2017)\citenamefont
  {Genevet}, \citenamefont {Capasso}, \citenamefont {Aieta}, \citenamefont
  {Khorasaninejad},\ and\ \citenamefont {Devlin}}]{genevet2017recent}%
  \BibitemOpen
  \bibfield  {author} {\bibinfo {author} {\bibfnamefont {Patrice}\ \bibnamefont
  {Genevet}}, \bibinfo {author} {\bibfnamefont {Federico}\ \bibnamefont
  {Capasso}}, \bibinfo {author} {\bibfnamefont {Francesco}\ \bibnamefont
  {Aieta}}, \bibinfo {author} {\bibfnamefont {Mohammadreza}\ \bibnamefont
  {Khorasaninejad}}, \ and\ \bibinfo {author} {\bibfnamefont {Robert}\
  \bibnamefont {Devlin}},\ }\bibfield  {title} {\enquote {\bibinfo {title}
  {Recent advances in planar optics: from plasmonic to dielectric
  metasurfaces},}\ }\href@noop {} {\bibfield  {journal} {\bibinfo  {journal}
  {Optica}\ }\textbf {\bibinfo {volume} {4}},\ \bibinfo {pages} {139--152}
  (\bibinfo {year} {2017})}\BibitemShut {NoStop}%
\bibitem [{\citenamefont {Banerji}\ \emph {et~al.}(2019)\citenamefont
  {Banerji}, \citenamefont {Meem}, \citenamefont {Majumder}, \citenamefont
  {Vasquez}, \citenamefont {Sensale-Rodriguez},\ and\ \citenamefont
  {Menon}}]{banerji2019imaging}%
  \BibitemOpen
  \bibfield  {author} {\bibinfo {author} {\bibfnamefont {Sourangsu}\
  \bibnamefont {Banerji}}, \bibinfo {author} {\bibfnamefont {Monjurul}\
  \bibnamefont {Meem}}, \bibinfo {author} {\bibfnamefont {Apratim}\
  \bibnamefont {Majumder}}, \bibinfo {author} {\bibfnamefont
  {Fernando~Guevara}\ \bibnamefont {Vasquez}}, \bibinfo {author} {\bibfnamefont
  {Berardi}\ \bibnamefont {Sensale-Rodriguez}}, \ and\ \bibinfo {author}
  {\bibfnamefont {Rajesh}\ \bibnamefont {Menon}},\ }\bibfield  {title}
  {\enquote {\bibinfo {title} {Imaging with flat optics: metalenses or
  diffractive lenses?}}\ }\href@noop {} {\bibfield  {journal} {\bibinfo
  {journal} {Optica}\ }\textbf {\bibinfo {volume} {6}},\ \bibinfo {pages}
  {805--810} (\bibinfo {year} {2019})}\BibitemShut {NoStop}%
\bibitem [{\citenamefont {Hu}\ \emph {et~al.}(2016)\citenamefont {Hu},
  \citenamefont {Xu}, \citenamefont {Wen}, \citenamefont {Wang}, \citenamefont
  {Zhao}, \citenamefont {Zhang}, \citenamefont {Cumming},\ and\ \citenamefont
  {Chen}}]{hu2016metamaterial}%
  \BibitemOpen
  \bibfield  {author} {\bibinfo {author} {\bibfnamefont {Xin}\ \bibnamefont
  {Hu}}, \bibinfo {author} {\bibfnamefont {Gaiqi}\ \bibnamefont {Xu}}, \bibinfo
  {author} {\bibfnamefont {Long}\ \bibnamefont {Wen}}, \bibinfo {author}
  {\bibfnamefont {Huacun}\ \bibnamefont {Wang}}, \bibinfo {author}
  {\bibfnamefont {Yuncheng}\ \bibnamefont {Zhao}}, \bibinfo {author}
  {\bibfnamefont {Yaxin}\ \bibnamefont {Zhang}}, \bibinfo {author}
  {\bibfnamefont {David~RS}\ \bibnamefont {Cumming}}, \ and\ \bibinfo {author}
  {\bibfnamefont {Qin}\ \bibnamefont {Chen}},\ }\bibfield  {title} {\enquote
  {\bibinfo {title} {Metamaterial absorber integrated microfluidic terahertz
  sensors},}\ }\href@noop {} {\bibfield  {journal} {\bibinfo  {journal} {Laser
  \& Photonics Reviews}\ }\textbf {\bibinfo {volume} {10}},\ \bibinfo {pages}
  {962--969} (\bibinfo {year} {2016})}\BibitemShut {NoStop}%
\bibitem [{\citenamefont {Cheben}\ \emph {et~al.}(2018)\citenamefont {Cheben},
  \citenamefont {Halir}, \citenamefont {Schmid}, \citenamefont {Atwater},\ and\
  \citenamefont {Smith}}]{cheben2018subwavelength}%
  \BibitemOpen
  \bibfield  {author} {\bibinfo {author} {\bibfnamefont {Pavel}\ \bibnamefont
  {Cheben}}, \bibinfo {author} {\bibfnamefont {Robert}\ \bibnamefont {Halir}},
  \bibinfo {author} {\bibfnamefont {Jens~H}\ \bibnamefont {Schmid}}, \bibinfo
  {author} {\bibfnamefont {Harry~A}\ \bibnamefont {Atwater}}, \ and\ \bibinfo
  {author} {\bibfnamefont {David~R}\ \bibnamefont {Smith}},\ }\bibfield
  {title} {\enquote {\bibinfo {title} {Subwavelength integrated photonics},}\
  }\href@noop {} {\bibfield  {journal} {\bibinfo  {journal} {Nature}\ }\textbf
  {\bibinfo {volume} {560}},\ \bibinfo {pages} {565--572} (\bibinfo {year}
  {2018})}\BibitemShut {NoStop}%
\bibitem [{\citenamefont {Romagnoli}\ \emph {et~al.}(2018)\citenamefont
  {Romagnoli}, \citenamefont {Sorianello}, \citenamefont {Midrio},
  \citenamefont {Koppens}, \citenamefont {Huyghebaert}, \citenamefont
  {Neumaier}, \citenamefont {Galli}, \citenamefont {Templ}, \citenamefont
  {D'Errico},\ and\ \citenamefont {Ferrari}}]{romagnoli2018graphene}%
  \BibitemOpen
  \bibfield  {author} {\bibinfo {author} {\bibfnamefont {Marco}\ \bibnamefont
  {Romagnoli}}, \bibinfo {author} {\bibfnamefont {Vito}\ \bibnamefont
  {Sorianello}}, \bibinfo {author} {\bibfnamefont {Michele}\ \bibnamefont
  {Midrio}}, \bibinfo {author} {\bibfnamefont {Frank~HL}\ \bibnamefont
  {Koppens}}, \bibinfo {author} {\bibfnamefont {Cedric}\ \bibnamefont
  {Huyghebaert}}, \bibinfo {author} {\bibfnamefont {Daniel}\ \bibnamefont
  {Neumaier}}, \bibinfo {author} {\bibfnamefont {Paola}\ \bibnamefont {Galli}},
  \bibinfo {author} {\bibfnamefont {Wolfgang}\ \bibnamefont {Templ}}, \bibinfo
  {author} {\bibfnamefont {Antonio}\ \bibnamefont {D'Errico}}, \ and\ \bibinfo
  {author} {\bibfnamefont {Andrea~C}\ \bibnamefont {Ferrari}},\ }\bibfield
  {title} {\enquote {\bibinfo {title} {Graphene-based integrated photonics for
  next-generation datacom and telecom},}\ }\href@noop {} {\bibfield  {journal}
  {\bibinfo  {journal} {Nature Reviews Materials}\ }\textbf {\bibinfo {volume}
  {3}},\ \bibinfo {pages} {392--414} (\bibinfo {year} {2018})}\BibitemShut
  {NoStop}%
\bibitem [{\citenamefont {Redding}\ \emph {et~al.}(2013)\citenamefont
  {Redding}, \citenamefont {Liew}, \citenamefont {Sarma},\ and\ \citenamefont
  {Cao}}]{redding2013compact}%
  \BibitemOpen
  \bibfield  {author} {\bibinfo {author} {\bibfnamefont {Brandon}\ \bibnamefont
  {Redding}}, \bibinfo {author} {\bibfnamefont {Seng~Fatt}\ \bibnamefont
  {Liew}}, \bibinfo {author} {\bibfnamefont {Raktim}\ \bibnamefont {Sarma}}, \
  and\ \bibinfo {author} {\bibfnamefont {Hui}\ \bibnamefont {Cao}},\ }\bibfield
   {title} {\enquote {\bibinfo {title} {Compact spectrometer based on a
  disordered photonic chip},}\ }\href@noop {} {\bibfield  {journal} {\bibinfo
  {journal} {Nature Photonics}\ }\textbf {\bibinfo {volume} {7}},\ \bibinfo
  {pages} {746} (\bibinfo {year} {2013})}\BibitemShut {NoStop}%
\bibitem [{\citenamefont {Boschetti}\ \emph {et~al.}(2019)\citenamefont
  {Boschetti}, \citenamefont {Taschin}, \citenamefont {Bartolini},
  \citenamefont {Tiwari}, \citenamefont {Pattelli}, \citenamefont {Torre},\
  and\ \citenamefont {Wiersma}}]{boschetti2019spectral}%
  \BibitemOpen
  \bibfield  {author} {\bibinfo {author} {\bibfnamefont {Alice}\ \bibnamefont
  {Boschetti}}, \bibinfo {author} {\bibfnamefont {Andrea}\ \bibnamefont
  {Taschin}}, \bibinfo {author} {\bibfnamefont {Paolo}\ \bibnamefont
  {Bartolini}}, \bibinfo {author} {\bibfnamefont {Anjani~Kumar}\ \bibnamefont
  {Tiwari}}, \bibinfo {author} {\bibfnamefont {Lorenzo}\ \bibnamefont
  {Pattelli}}, \bibinfo {author} {\bibfnamefont {Renato}\ \bibnamefont
  {Torre}}, \ and\ \bibinfo {author} {\bibfnamefont {Diederik~S}\ \bibnamefont
  {Wiersma}},\ }\bibfield  {title} {\enquote {\bibinfo {title} {Spectral
  super-resolution spectroscopy using a random laser},}\ }\href@noop {}
  {\bibfield  {journal} {\bibinfo  {journal} {Nature Photonics}\ ,\ \bibinfo
  {pages} {1--6}} (\bibinfo {year} {2019})}\BibitemShut {NoStop}%
\bibitem [{\citenamefont {Kildishev}\ \emph {et~al.}(2013)\citenamefont
  {Kildishev}, \citenamefont {Boltasseva},\ and\ \citenamefont
  {Shalaev}}]{kildishev2013planar}%
  \BibitemOpen
  \bibfield  {author} {\bibinfo {author} {\bibfnamefont {Alexander~V}\
  \bibnamefont {Kildishev}}, \bibinfo {author} {\bibfnamefont {Alexandra}\
  \bibnamefont {Boltasseva}}, \ and\ \bibinfo {author} {\bibfnamefont
  {Vladimir~M}\ \bibnamefont {Shalaev}},\ }\bibfield  {title} {\enquote
  {\bibinfo {title} {Planar photonics with metasurfaces},}\ }\href@noop {}
  {\bibfield  {journal} {\bibinfo  {journal} {Science}\ }\textbf {\bibinfo
  {volume} {339}},\ \bibinfo {pages} {1232009} (\bibinfo {year}
  {2013})}\BibitemShut {NoStop}%
\bibitem [{\citenamefont {Koenderink}\ \emph {et~al.}(2015)\citenamefont
  {Koenderink}, \citenamefont {Alu},\ and\ \citenamefont
  {Polman}}]{koenderink2015nanophotonics}%
  \BibitemOpen
  \bibfield  {author} {\bibinfo {author} {\bibfnamefont {A~Femius}\
  \bibnamefont {Koenderink}}, \bibinfo {author} {\bibfnamefont {Andrea}\
  \bibnamefont {Alu}}, \ and\ \bibinfo {author} {\bibfnamefont {Albert}\
  \bibnamefont {Polman}},\ }\bibfield  {title} {\enquote {\bibinfo {title}
  {Nanophotonics: shrinking light-based technology},}\ }\href@noop {}
  {\bibfield  {journal} {\bibinfo  {journal} {Science}\ }\textbf {\bibinfo
  {volume} {348}},\ \bibinfo {pages} {516--521} (\bibinfo {year}
  {2015})}\BibitemShut {NoStop}%
\bibitem [{\citenamefont {Zheludev}\ and\ \citenamefont
  {Kivshar}(2012)}]{zheludev2012metamaterials}%
  \BibitemOpen
  \bibfield  {author} {\bibinfo {author} {\bibfnamefont {Nikolay~I}\
  \bibnamefont {Zheludev}}\ and\ \bibinfo {author} {\bibfnamefont {Yuri~S}\
  \bibnamefont {Kivshar}},\ }\bibfield  {title} {\enquote {\bibinfo {title}
  {From metamaterials to metadevices},}\ }\href@noop {} {\bibfield  {journal}
  {\bibinfo  {journal} {Nature materials}\ }\textbf {\bibinfo {volume} {11}},\
  \bibinfo {pages} {917--924} (\bibinfo {year} {2012})}\BibitemShut {NoStop}%
\bibitem [{\citenamefont {Yu}\ \emph {et~al.}(2011)\citenamefont {Yu},
  \citenamefont {Genevet}, \citenamefont {Kats}, \citenamefont {Aieta},
  \citenamefont {Tetienne}, \citenamefont {Capasso},\ and\ \citenamefont
  {Gaburro}}]{yu2011light}%
  \BibitemOpen
  \bibfield  {author} {\bibinfo {author} {\bibfnamefont {Nanfang}\ \bibnamefont
  {Yu}}, \bibinfo {author} {\bibfnamefont {Patrice}\ \bibnamefont {Genevet}},
  \bibinfo {author} {\bibfnamefont {Mikhail~A}\ \bibnamefont {Kats}}, \bibinfo
  {author} {\bibfnamefont {Francesco}\ \bibnamefont {Aieta}}, \bibinfo {author}
  {\bibfnamefont {Jean-Philippe}\ \bibnamefont {Tetienne}}, \bibinfo {author}
  {\bibfnamefont {Federico}\ \bibnamefont {Capasso}}, \ and\ \bibinfo {author}
  {\bibfnamefont {Zeno}\ \bibnamefont {Gaburro}},\ }\bibfield  {title}
  {\enquote {\bibinfo {title} {Light propagation with phase discontinuities:
  generalized laws of reflection and refraction},}\ }\href@noop {} {\bibfield
  {journal} {\bibinfo  {journal} {science}\ }\textbf {\bibinfo {volume}
  {334}},\ \bibinfo {pages} {333--337} (\bibinfo {year} {2011})}\BibitemShut
  {NoStop}%
\bibitem [{\citenamefont {Yu}\ and\ \citenamefont
  {Capasso}(2014)}]{yu2014flat}%
  \BibitemOpen
  \bibfield  {author} {\bibinfo {author} {\bibfnamefont {Nanfang}\ \bibnamefont
  {Yu}}\ and\ \bibinfo {author} {\bibfnamefont {Federico}\ \bibnamefont
  {Capasso}},\ }\bibfield  {title} {\enquote {\bibinfo {title} {Flat optics
  with designer metasurfaces},}\ }\href@noop {} {\bibfield  {journal} {\bibinfo
   {journal} {Nature materials}\ }\textbf {\bibinfo {volume} {13}},\ \bibinfo
  {pages} {139--150} (\bibinfo {year} {2014})}\BibitemShut {NoStop}%
\bibitem [{\citenamefont {Zhang}\ and\ \citenamefont
  {Liu}(2008)}]{zhang2008superlenses}%
  \BibitemOpen
  \bibfield  {author} {\bibinfo {author} {\bibfnamefont {Xiang}\ \bibnamefont
  {Zhang}}\ and\ \bibinfo {author} {\bibfnamefont {Zhaowei}\ \bibnamefont
  {Liu}},\ }\bibfield  {title} {\enquote {\bibinfo {title} {Superlenses to
  overcome the diffraction limit},}\ }\href@noop {} {\bibfield  {journal}
  {\bibinfo  {journal} {Nature materials}\ }\textbf {\bibinfo {volume} {7}},\
  \bibinfo {pages} {435--441} (\bibinfo {year} {2008})}\BibitemShut {NoStop}%
\bibitem [{\citenamefont {Rogers}\ \emph {et~al.}(2012)\citenamefont {Rogers},
  \citenamefont {Lindberg}, \citenamefont {Roy}, \citenamefont {Savo},
  \citenamefont {Chad}, \citenamefont {Dennis},\ and\ \citenamefont
  {Zheludev}}]{rogers2012super}%
  \BibitemOpen
  \bibfield  {author} {\bibinfo {author} {\bibfnamefont {Edward~TF}\
  \bibnamefont {Rogers}}, \bibinfo {author} {\bibfnamefont {Jari}\ \bibnamefont
  {Lindberg}}, \bibinfo {author} {\bibfnamefont {Tapashree}\ \bibnamefont
  {Roy}}, \bibinfo {author} {\bibfnamefont {Salvatore}\ \bibnamefont {Savo}},
  \bibinfo {author} {\bibfnamefont {John~E}\ \bibnamefont {Chad}}, \bibinfo
  {author} {\bibfnamefont {Mark~R}\ \bibnamefont {Dennis}}, \ and\ \bibinfo
  {author} {\bibfnamefont {Nikolay~I}\ \bibnamefont {Zheludev}},\ }\bibfield
  {title} {\enquote {\bibinfo {title} {A super-oscillatory lens optical
  microscope for subwavelength imaging},}\ }\href@noop {} {\bibfield  {journal}
  {\bibinfo  {journal} {Nature materials}\ }\textbf {\bibinfo {volume} {11}},\
  \bibinfo {pages} {432--435} (\bibinfo {year} {2012})}\BibitemShut {NoStop}%
\bibitem [{\citenamefont {Lalanne}\ and\ \citenamefont
  {Chavel}(2017)}]{lalanne2017metalenses}%
  \BibitemOpen
  \bibfield  {author} {\bibinfo {author} {\bibfnamefont {Philippe}\
  \bibnamefont {Lalanne}}\ and\ \bibinfo {author} {\bibfnamefont {Pierre}\
  \bibnamefont {Chavel}},\ }\bibfield  {title} {\enquote {\bibinfo {title}
  {Metalenses at visible wavelengths: past, present, perspectives},}\
  }\href@noop {} {\bibfield  {journal} {\bibinfo  {journal} {Laser \& Photonics
  Reviews}\ }\textbf {\bibinfo {volume} {11}},\ \bibinfo {pages} {1600295}
  (\bibinfo {year} {2017})}\BibitemShut {NoStop}%
\bibitem [{\citenamefont {Zhu}\ \emph {et~al.}(2017)\citenamefont {Zhu},
  \citenamefont {Kuznetsov}, \citenamefont {Luk'yanchuk}, \citenamefont
  {Engheta},\ and\ \citenamefont {Genevet}}]{zhu2017traditional}%
  \BibitemOpen
  \bibfield  {author} {\bibinfo {author} {\bibfnamefont {Alexander~Y}\
  \bibnamefont {Zhu}}, \bibinfo {author} {\bibfnamefont {Arseniy~I}\
  \bibnamefont {Kuznetsov}}, \bibinfo {author} {\bibfnamefont {Boris}\
  \bibnamefont {Luk'yanchuk}}, \bibinfo {author} {\bibfnamefont {Nader}\
  \bibnamefont {Engheta}}, \ and\ \bibinfo {author} {\bibfnamefont {Patrice}\
  \bibnamefont {Genevet}},\ }\bibfield  {title} {\enquote {\bibinfo {title}
  {Traditional and emerging materials for optical metasurfaces},}\ }\href@noop
  {} {\bibfield  {journal} {\bibinfo  {journal} {Nanophotonics}\ }\textbf
  {\bibinfo {volume} {6}},\ \bibinfo {pages} {452--471} (\bibinfo {year}
  {2017})}\BibitemShut {NoStop}%
\bibitem [{\citenamefont {Kuznetsov}\ \emph {et~al.}(2016)\citenamefont
  {Kuznetsov}, \citenamefont {Miroshnichenko}, \citenamefont {Brongersma},
  \citenamefont {Kivshar},\ and\ \citenamefont
  {Luk'yanchuk}}]{kuznetsov2016optically}%
  \BibitemOpen
  \bibfield  {author} {\bibinfo {author} {\bibfnamefont {Arseniy~I}\
  \bibnamefont {Kuznetsov}}, \bibinfo {author} {\bibfnamefont {Andrey~E}\
  \bibnamefont {Miroshnichenko}}, \bibinfo {author} {\bibfnamefont {Mark~L}\
  \bibnamefont {Brongersma}}, \bibinfo {author} {\bibfnamefont {Yuri~S}\
  \bibnamefont {Kivshar}}, \ and\ \bibinfo {author} {\bibfnamefont {Boris}\
  \bibnamefont {Luk'yanchuk}},\ }\bibfield  {title} {\enquote {\bibinfo {title}
  {Optically resonant dielectric nanostructures},}\ }\href@noop {} {\bibfield
  {journal} {\bibinfo  {journal} {Science}\ }\textbf {\bibinfo {volume}
  {354}},\ \bibinfo {pages} {aag2472} (\bibinfo {year} {2016})}\BibitemShut
  {NoStop}%
\bibitem [{\citenamefont {Arbabi}\ \emph {et~al.}(2015)\citenamefont {Arbabi},
  \citenamefont {Horie}, \citenamefont {Ball}, \citenamefont {Bagheri},\ and\
  \citenamefont {Faraon}}]{arbabi2015subwavelength}%
  \BibitemOpen
  \bibfield  {author} {\bibinfo {author} {\bibfnamefont {Amir}\ \bibnamefont
  {Arbabi}}, \bibinfo {author} {\bibfnamefont {Yu}~\bibnamefont {Horie}},
  \bibinfo {author} {\bibfnamefont {Alexander~J}\ \bibnamefont {Ball}},
  \bibinfo {author} {\bibfnamefont {Mahmood}\ \bibnamefont {Bagheri}}, \ and\
  \bibinfo {author} {\bibfnamefont {Andrei}\ \bibnamefont {Faraon}},\
  }\bibfield  {title} {\enquote {\bibinfo {title} {Subwavelength-thick lenses
  with high numerical apertures and large efficiency based on high-contrast
  transmitarrays},}\ }\href@noop {} {\bibfield  {journal} {\bibinfo  {journal}
  {Nature Communications}\ }\textbf {\bibinfo {volume} {6}},\ \bibinfo {pages}
  {7069} (\bibinfo {year} {2015})}\BibitemShut {NoStop}%
\bibitem [{\citenamefont {Khorasaninejad}\ \emph
  {et~al.}(2016{\natexlab{a}})\citenamefont {Khorasaninejad}, \citenamefont
  {Zhu}, \citenamefont {Roques-Carmes}, \citenamefont {Chen}, \citenamefont
  {Oh}, \citenamefont {Mishra}, \citenamefont {Devlin},\ and\ \citenamefont
  {Capasso}}]{khorasaninejad2016polarization}%
  \BibitemOpen
  \bibfield  {author} {\bibinfo {author} {\bibfnamefont {Mohammadreza}\
  \bibnamefont {Khorasaninejad}}, \bibinfo {author} {\bibfnamefont
  {Alexander~Yutong}\ \bibnamefont {Zhu}}, \bibinfo {author} {\bibfnamefont
  {Charles}\ \bibnamefont {Roques-Carmes}}, \bibinfo {author} {\bibfnamefont
  {Wei~Ting}\ \bibnamefont {Chen}}, \bibinfo {author} {\bibfnamefont {Jaewon}\
  \bibnamefont {Oh}}, \bibinfo {author} {\bibfnamefont {Ishan}\ \bibnamefont
  {Mishra}}, \bibinfo {author} {\bibfnamefont {Robert~C}\ \bibnamefont
  {Devlin}}, \ and\ \bibinfo {author} {\bibfnamefont {Federico}\ \bibnamefont
  {Capasso}},\ }\bibfield  {title} {\enquote {\bibinfo {title}
  {Polarization-insensitive metalenses at visible wavelengths},}\ }\href@noop
  {} {\bibfield  {journal} {\bibinfo  {journal} {Nano Letters}\ }\textbf
  {\bibinfo {volume} {16}},\ \bibinfo {pages} {7229--7234} (\bibinfo {year}
  {2016}{\natexlab{a}})}\BibitemShut {NoStop}%
\bibitem [{\citenamefont {Chen}\ \emph {et~al.}(2019)\citenamefont {Chen},
  \citenamefont {Zhu}, \citenamefont {Sisler}, \citenamefont {Bharwani},\ and\
  \citenamefont {Capasso}}]{chen2019broadband}%
  \BibitemOpen
  \bibfield  {author} {\bibinfo {author} {\bibfnamefont {Wei~Ting}\
  \bibnamefont {Chen}}, \bibinfo {author} {\bibfnamefont {Alexander~Y}\
  \bibnamefont {Zhu}}, \bibinfo {author} {\bibfnamefont {Jared}\ \bibnamefont
  {Sisler}}, \bibinfo {author} {\bibfnamefont {Zameer}\ \bibnamefont
  {Bharwani}}, \ and\ \bibinfo {author} {\bibfnamefont {Federico}\ \bibnamefont
  {Capasso}},\ }\bibfield  {title} {\enquote {\bibinfo {title} {A broadband
  achromatic polarization-insensitive metalens consisting of anisotropic
  nanostructures},}\ }\href@noop {} {\bibfield  {journal} {\bibinfo  {journal}
  {Nature Communications}\ }\textbf {\bibinfo {volume} {10}},\ \bibinfo {pages}
  {355} (\bibinfo {year} {2019})}\BibitemShut {NoStop}%
\bibitem [{\citenamefont {Lin}\ \emph {et~al.}(2014)\citenamefont {Lin},
  \citenamefont {Fan}, \citenamefont {Hasman},\ and\ \citenamefont
  {Brongersma}}]{lin2014dielectric}%
  \BibitemOpen
  \bibfield  {author} {\bibinfo {author} {\bibfnamefont {Dianmin}\ \bibnamefont
  {Lin}}, \bibinfo {author} {\bibfnamefont {Pengyu}\ \bibnamefont {Fan}},
  \bibinfo {author} {\bibfnamefont {Erez}\ \bibnamefont {Hasman}}, \ and\
  \bibinfo {author} {\bibfnamefont {Mark~L}\ \bibnamefont {Brongersma}},\
  }\bibfield  {title} {\enquote {\bibinfo {title} {Dielectric gradient
  metasurface optical elements},}\ }\href@noop {} {\bibfield  {journal}
  {\bibinfo  {journal} {Science}\ }\textbf {\bibinfo {volume} {345}},\ \bibinfo
  {pages} {298--302} (\bibinfo {year} {2014})}\BibitemShut {NoStop}%
\bibitem [{\citenamefont {Zheng}\ \emph {et~al.}(2015)\citenamefont {Zheng},
  \citenamefont {M{\"u}hlenbernd}, \citenamefont {Kenney}, \citenamefont {Li},
  \citenamefont {Zentgraf},\ and\ \citenamefont
  {Zhang}}]{zheng2015metasurface}%
  \BibitemOpen
  \bibfield  {author} {\bibinfo {author} {\bibfnamefont {Guoxing}\ \bibnamefont
  {Zheng}}, \bibinfo {author} {\bibfnamefont {Holger}\ \bibnamefont
  {M{\"u}hlenbernd}}, \bibinfo {author} {\bibfnamefont {Mitchell}\ \bibnamefont
  {Kenney}}, \bibinfo {author} {\bibfnamefont {Guixin}\ \bibnamefont {Li}},
  \bibinfo {author} {\bibfnamefont {Thomas}\ \bibnamefont {Zentgraf}}, \ and\
  \bibinfo {author} {\bibfnamefont {Shuang}\ \bibnamefont {Zhang}},\ }\bibfield
   {title} {\enquote {\bibinfo {title} {Metasurface holograms reaching 80\%
  efficiency},}\ }\href@noop {} {\bibfield  {journal} {\bibinfo  {journal}
  {Nature Nanotechnology}\ }\textbf {\bibinfo {volume} {10}},\ \bibinfo {pages}
  {308} (\bibinfo {year} {2015})}\BibitemShut {NoStop}%
\bibitem [{\citenamefont {Zhan}\ \emph {et~al.}(2016)\citenamefont {Zhan},
  \citenamefont {Colburn}, \citenamefont {Trivedi}, \citenamefont {Fryett},
  \citenamefont {Dodson},\ and\ \citenamefont {Majumdar}}]{zhan2016low}%
  \BibitemOpen
  \bibfield  {author} {\bibinfo {author} {\bibfnamefont {Alan}\ \bibnamefont
  {Zhan}}, \bibinfo {author} {\bibfnamefont {Shane}\ \bibnamefont {Colburn}},
  \bibinfo {author} {\bibfnamefont {Rahul}\ \bibnamefont {Trivedi}}, \bibinfo
  {author} {\bibfnamefont {Taylor~K}\ \bibnamefont {Fryett}}, \bibinfo {author}
  {\bibfnamefont {Christopher~M}\ \bibnamefont {Dodson}}, \ and\ \bibinfo
  {author} {\bibfnamefont {Arka}\ \bibnamefont {Majumdar}},\ }\bibfield
  {title} {\enquote {\bibinfo {title} {Low-contrast dielectric metasurface
  optics},}\ }\href@noop {} {\bibfield  {journal} {\bibinfo  {journal} {ACS
  Photonics}\ }\textbf {\bibinfo {volume} {3}},\ \bibinfo {pages} {209--214}
  (\bibinfo {year} {2016})}\BibitemShut {NoStop}%
\bibitem [{\citenamefont {Fleming}\ and\ \citenamefont
  {Hutley}(1997)}]{fleming1997blazed}%
  \BibitemOpen
  \bibfield  {author} {\bibinfo {author} {\bibfnamefont {Madeleine~B}\
  \bibnamefont {Fleming}}\ and\ \bibinfo {author} {\bibfnamefont
  {MC}~\bibnamefont {Hutley}},\ }\bibfield  {title} {\enquote {\bibinfo {title}
  {Blazed diffractive optics},}\ }\href@noop {} {\bibfield  {journal} {\bibinfo
   {journal} {Applied optics}\ }\textbf {\bibinfo {volume} {36}},\ \bibinfo
  {pages} {4635--4643} (\bibinfo {year} {1997})}\BibitemShut {NoStop}%
\bibitem [{\citenamefont {Swanson}(1989)}]{swanson1989binary}%
  \BibitemOpen
  \bibfield  {author} {\bibinfo {author} {\bibfnamefont {Gary~J}\ \bibnamefont
  {Swanson}},\ }\href@noop {} {\emph {\bibinfo {title} {Binary optics
  technology: the theory and design of multi-level diffractive optical
  elements}}},\ \bibinfo {type} {Tech. Rep.}\ (\bibinfo  {institution}
  {MASSACHUSETTS INST OF TECH LEXINGTON LINCOLN LAB},\ \bibinfo {year}
  {1989})\BibitemShut {NoStop}%
\bibitem [{\citenamefont {Mohammad}\ \emph {et~al.}(2018)\citenamefont
  {Mohammad}, \citenamefont {Meem}, \citenamefont {Shen}, \citenamefont
  {Wang},\ and\ \citenamefont {Menon}}]{mohammad2018broadband}%
  \BibitemOpen
  \bibfield  {author} {\bibinfo {author} {\bibfnamefont {Nabil}\ \bibnamefont
  {Mohammad}}, \bibinfo {author} {\bibfnamefont {Monjurul}\ \bibnamefont
  {Meem}}, \bibinfo {author} {\bibfnamefont {Bing}\ \bibnamefont {Shen}},
  \bibinfo {author} {\bibfnamefont {Peng}\ \bibnamefont {Wang}}, \ and\
  \bibinfo {author} {\bibfnamefont {Rajesh}\ \bibnamefont {Menon}},\ }\bibfield
   {title} {\enquote {\bibinfo {title} {Broadband imaging with one planar
  diffractive lens},}\ }\href@noop {} {\bibfield  {journal} {\bibinfo
  {journal} {Scientific reports}\ }\textbf {\bibinfo {volume} {8}},\ \bibinfo
  {pages} {1--6} (\bibinfo {year} {2018})}\BibitemShut {NoStop}%
\bibitem [{\citenamefont {Chao}\ \emph {et~al.}(2005)\citenamefont {Chao},
  \citenamefont {Patel}, \citenamefont {Barwicz}, \citenamefont {Smith},\ and\
  \citenamefont {Menon}}]{chao2005immersion}%
  \BibitemOpen
  \bibfield  {author} {\bibinfo {author} {\bibfnamefont {David}\ \bibnamefont
  {Chao}}, \bibinfo {author} {\bibfnamefont {Amil}\ \bibnamefont {Patel}},
  \bibinfo {author} {\bibfnamefont {Tymon}\ \bibnamefont {Barwicz}}, \bibinfo
  {author} {\bibfnamefont {Henry~I}\ \bibnamefont {Smith}}, \ and\ \bibinfo
  {author} {\bibfnamefont {Rajesh}\ \bibnamefont {Menon}},\ }\bibfield  {title}
  {\enquote {\bibinfo {title} {Immersion zone-plate-array lithography},}\
  }\href@noop {} {\bibfield  {journal} {\bibinfo  {journal} {Journal of Vacuum
  Science \& Technology B: Microelectronics and Nanometer Structures
  Processing, Measurement, and Phenomena}\ }\textbf {\bibinfo {volume} {23}},\
  \bibinfo {pages} {2657--2661} (\bibinfo {year} {2005})}\BibitemShut {NoStop}%
\bibitem [{\citenamefont {Davidson}\ \emph {et~al.}(1991)\citenamefont
  {Davidson}, \citenamefont {Friesem},\ and\ \citenamefont
  {Hasman}}]{davidson1991holographic}%
  \BibitemOpen
  \bibfield  {author} {\bibinfo {author} {\bibfnamefont {N}~\bibnamefont
  {Davidson}}, \bibinfo {author} {\bibfnamefont {AA}~\bibnamefont {Friesem}}, \
  and\ \bibinfo {author} {\bibfnamefont {E}~\bibnamefont {Hasman}},\ }\bibfield
   {title} {\enquote {\bibinfo {title} {Holographic axilens: high resolution
  and long focal depth},}\ }\href@noop {} {\bibfield  {journal} {\bibinfo
  {journal} {Optics letters}\ }\textbf {\bibinfo {volume} {16}},\ \bibinfo
  {pages} {523--525} (\bibinfo {year} {1991})}\BibitemShut {NoStop}%
\bibitem [{\citenamefont {Tang}\ \emph {et~al.}(2019)\citenamefont {Tang},
  \citenamefont {Ackerman}, \citenamefont {Chen},\ and\ \citenamefont
  {Guyot-Sionnest}}]{tang2019dual}%
  \BibitemOpen
  \bibfield  {author} {\bibinfo {author} {\bibfnamefont {Xin}\ \bibnamefont
  {Tang}}, \bibinfo {author} {\bibfnamefont {Matthew~M}\ \bibnamefont
  {Ackerman}}, \bibinfo {author} {\bibfnamefont {Menglu}\ \bibnamefont {Chen}},
  \ and\ \bibinfo {author} {\bibfnamefont {Philippe}\ \bibnamefont
  {Guyot-Sionnest}},\ }\bibfield  {title} {\enquote {\bibinfo {title}
  {Dual-band infrared imaging using stacked colloidal quantum dot
  photodiodes},}\ }\href@noop {} {\bibfield  {journal} {\bibinfo  {journal}
  {Nature Photonics}\ }\textbf {\bibinfo {volume} {13}},\ \bibinfo {pages}
  {277} (\bibinfo {year} {2019})}\BibitemShut {NoStop}%
\bibitem [{\citenamefont {{Gunapala}}\ \emph {et~al.}(2010)\citenamefont
  {{Gunapala}}, \citenamefont {{Bandara}}, \citenamefont {{Liu}}, \citenamefont
  {{Mumolo}}, \citenamefont {{Ting}}, \citenamefont {{Hill}}, \citenamefont
  {{Nguyen}}, \citenamefont {{Simolon}}, \citenamefont {{Woolaway}},
  \citenamefont {{Wang}}, \citenamefont {{Li}}, \citenamefont {{LeVan}},\ and\
  \citenamefont {{Tidrow}}}]{5401117}%
  \BibitemOpen
  \bibfield  {author} {\bibinfo {author} {\bibfnamefont {S.~D.}\ \bibnamefont
  {{Gunapala}}}, \bibinfo {author} {\bibfnamefont {S.~V.}\ \bibnamefont
  {{Bandara}}}, \bibinfo {author} {\bibfnamefont {J.~K.}\ \bibnamefont
  {{Liu}}}, \bibinfo {author} {\bibfnamefont {J.~M.}\ \bibnamefont {{Mumolo}}},
  \bibinfo {author} {\bibfnamefont {D.~Z.}\ \bibnamefont {{Ting}}}, \bibinfo
  {author} {\bibfnamefont {C.~J.}\ \bibnamefont {{Hill}}}, \bibinfo {author}
  {\bibfnamefont {J.}~\bibnamefont {{Nguyen}}}, \bibinfo {author}
  {\bibfnamefont {B.}~\bibnamefont {{Simolon}}}, \bibinfo {author}
  {\bibfnamefont {J.}~\bibnamefont {{Woolaway}}}, \bibinfo {author}
  {\bibfnamefont {S.~C.}\ \bibnamefont {{Wang}}}, \bibinfo {author}
  {\bibfnamefont {W.}~\bibnamefont {{Li}}}, \bibinfo {author} {\bibfnamefont
  {P.~D.}\ \bibnamefont {{LeVan}}}, \ and\ \bibinfo {author} {\bibfnamefont
  {M.~Z.}\ \bibnamefont {{Tidrow}}},\ }\bibfield  {title} {\enquote {\bibinfo
  {title} {Demonstration of megapixel dual-band qwip focal plane array},}\
  }\href@noop {} {\bibfield  {journal} {\bibinfo  {journal} {IEEE Journal of
  Quantum Electronics}\ }\textbf {\bibinfo {volume} {46}},\ \bibinfo {pages}
  {285--293} (\bibinfo {year} {2010})}\BibitemShut {NoStop}%
\bibitem [{\citenamefont {Lutz}\ \emph {et~al.}(2018)\citenamefont {Lutz},
  \citenamefont {Breiter}, \citenamefont {Eich}, \citenamefont {Figgemeier},
  \citenamefont {Hanna}, \citenamefont {Oelmaier},\ and\ \citenamefont
  {Wendler}}]{lutz2018towards}%
  \BibitemOpen
  \bibfield  {author} {\bibinfo {author} {\bibfnamefont {H}~\bibnamefont
  {Lutz}}, \bibinfo {author} {\bibfnamefont {R}~\bibnamefont {Breiter}},
  \bibinfo {author} {\bibfnamefont {D}~\bibnamefont {Eich}}, \bibinfo {author}
  {\bibfnamefont {H}~\bibnamefont {Figgemeier}}, \bibinfo {author}
  {\bibfnamefont {S}~\bibnamefont {Hanna}}, \bibinfo {author} {\bibfnamefont
  {R}~\bibnamefont {Oelmaier}}, \ and\ \bibinfo {author} {\bibfnamefont
  {J}~\bibnamefont {Wendler}},\ }\bibfield  {title} {\enquote {\bibinfo {title}
  {Towards ultra-small pixel pitch cooled mw and lw ir-modules},}\ }in\
  \href@noop {} {\emph {\bibinfo {booktitle} {Infrared Technology and
  Applications XLIV}}},\ Vol.\ \bibinfo {volume} {10624}\ (\bibinfo
  {organization} {International Society for Optics and Photonics},\ \bibinfo
  {year} {2018})\ p.\ \bibinfo {pages} {106240B}\BibitemShut {NoStop}%
\bibitem [{\citenamefont {Gunapala}\ \emph {et~al.}(2010)\citenamefont
  {Gunapala}, \citenamefont {Bandara}, \citenamefont {Liu}, \citenamefont
  {Mumolo}, \citenamefont {Ting}, \citenamefont {Hill}, \citenamefont {Nguyen},
  \citenamefont {Simolon}, \citenamefont {Woolaway}, \citenamefont {Wang} \emph
  {et~al.}}]{gunapala2010demonstration}%
  \BibitemOpen
  \bibfield  {author} {\bibinfo {author} {\bibfnamefont {Sarath~D}\
  \bibnamefont {Gunapala}}, \bibinfo {author} {\bibfnamefont {Sumith~V}\
  \bibnamefont {Bandara}}, \bibinfo {author} {\bibfnamefont {John~K}\
  \bibnamefont {Liu}}, \bibinfo {author} {\bibfnamefont {Jason~M}\ \bibnamefont
  {Mumolo}}, \bibinfo {author} {\bibfnamefont {David~Z}\ \bibnamefont {Ting}},
  \bibinfo {author} {\bibfnamefont {Cory~J}\ \bibnamefont {Hill}}, \bibinfo
  {author} {\bibfnamefont {Jean}\ \bibnamefont {Nguyen}}, \bibinfo {author}
  {\bibfnamefont {Brian}\ \bibnamefont {Simolon}}, \bibinfo {author}
  {\bibfnamefont {James}\ \bibnamefont {Woolaway}}, \bibinfo {author}
  {\bibfnamefont {Samuel~C}\ \bibnamefont {Wang}},  \emph {et~al.},\ }\bibfield
   {title} {\enquote {\bibinfo {title} {Demonstration of megapixel dual-band
  qwip focal plane array},}\ }\href@noop {} {\bibfield  {journal} {\bibinfo
  {journal} {IEEE Journal of Quantum Electronics}\ }\textbf {\bibinfo {volume}
  {46}},\ \bibinfo {pages} {285--293} (\bibinfo {year} {2010})}\BibitemShut
  {NoStop}%
\bibitem [{\citenamefont {Davies}\ \emph {et~al.}(1994)\citenamefont {Davies},
  \citenamefont {McCormick},\ and\ \citenamefont {Brewin}}]{davies1994design}%
  \BibitemOpen
  \bibfield  {author} {\bibinfo {author} {\bibfnamefont {Neil~A}\ \bibnamefont
  {Davies}}, \bibinfo {author} {\bibfnamefont {Malcolm}\ \bibnamefont
  {McCormick}}, \ and\ \bibinfo {author} {\bibfnamefont {Michael}\ \bibnamefont
  {Brewin}},\ }\bibfield  {title} {\enquote {\bibinfo {title} {Design and
  analysis of an image transfer system using microlens arrays},}\ }\href@noop
  {} {\bibfield  {journal} {\bibinfo  {journal} {Optical Engineering}\ }\textbf
  {\bibinfo {volume} {33}},\ \bibinfo {pages} {3624--3634} (\bibinfo {year}
  {1994})}\BibitemShut {NoStop}%
\bibitem [{\citenamefont {Nussbaum}\ \emph {et~al.}(1997)\citenamefont
  {Nussbaum}, \citenamefont {Voelkel}, \citenamefont {Herzig}, \citenamefont
  {Eisner},\ and\ \citenamefont {Haselbeck}}]{nussbaum1997design}%
  \BibitemOpen
  \bibfield  {author} {\bibinfo {author} {\bibfnamefont {Ph}~\bibnamefont
  {Nussbaum}}, \bibinfo {author} {\bibfnamefont {Reinhard}\ \bibnamefont
  {Voelkel}}, \bibinfo {author} {\bibfnamefont {Hans~Peter}\ \bibnamefont
  {Herzig}}, \bibinfo {author} {\bibfnamefont {Martin}\ \bibnamefont {Eisner}},
  \ and\ \bibinfo {author} {\bibfnamefont {Stefan}\ \bibnamefont {Haselbeck}},\
  }\bibfield  {title} {\enquote {\bibinfo {title} {Design, fabrication and
  testing of microlens arrays for sensors and microsystems},}\ }\href@noop {}
  {\bibfield  {journal} {\bibinfo  {journal} {Pure and applied optics: Journal
  of the European optical society part A}\ }\textbf {\bibinfo {volume} {6}},\
  \bibinfo {pages} {617} (\bibinfo {year} {1997})}\BibitemShut {NoStop}%
\bibitem [{\citenamefont {Chen}\ \emph {et~al.}(2015)\citenamefont {Chen},
  \citenamefont {Chen}, \citenamefont {Mehmood}, \citenamefont {Wen},
  \citenamefont {Yue}, \citenamefont {Qiu},\ and\ \citenamefont
  {Zhang}}]{chen2015longitudinal}%
  \BibitemOpen
  \bibfield  {author} {\bibinfo {author} {\bibfnamefont {Xianzhong}\
  \bibnamefont {Chen}}, \bibinfo {author} {\bibfnamefont {Ming}\ \bibnamefont
  {Chen}}, \bibinfo {author} {\bibfnamefont {Muhammad~Qasim}\ \bibnamefont
  {Mehmood}}, \bibinfo {author} {\bibfnamefont {Dandan}\ \bibnamefont {Wen}},
  \bibinfo {author} {\bibfnamefont {Fuyong}\ \bibnamefont {Yue}}, \bibinfo
  {author} {\bibfnamefont {Cheng-Wei}\ \bibnamefont {Qiu}}, \ and\ \bibinfo
  {author} {\bibfnamefont {Shuang}\ \bibnamefont {Zhang}},\ }\bibfield  {title}
  {\enquote {\bibinfo {title} {Longitudinal multifoci metalens for circularly
  polarized light},}\ }\href@noop {} {\bibfield  {journal} {\bibinfo  {journal}
  {Advanced Optical Materials}\ }\textbf {\bibinfo {volume} {3}},\ \bibinfo
  {pages} {1201--1206} (\bibinfo {year} {2015})}\BibitemShut {NoStop}%
\bibitem [{\citenamefont {Williams}\ \emph {et~al.}(2017)\citenamefont
  {Williams}, \citenamefont {Montelongo},\ and\ \citenamefont
  {Wilkinson}}]{williams2017plasmonic}%
  \BibitemOpen
  \bibfield  {author} {\bibinfo {author} {\bibfnamefont {Calum}\ \bibnamefont
  {Williams}}, \bibinfo {author} {\bibfnamefont {Yunuen}\ \bibnamefont
  {Montelongo}}, \ and\ \bibinfo {author} {\bibfnamefont {Timothy~D}\
  \bibnamefont {Wilkinson}},\ }\bibfield  {title} {\enquote {\bibinfo {title}
  {Plasmonic metalens for narrowband dual-focus imaging},}\ }\href@noop {}
  {\bibfield  {journal} {\bibinfo  {journal} {Advanced Optical Materials}\
  }\textbf {\bibinfo {volume} {5}},\ \bibinfo {pages} {1700811} (\bibinfo
  {year} {2017})}\BibitemShut {NoStop}%
\bibitem [{\citenamefont {Goodman}(2017)}]{goodman2005introduction}%
  \BibitemOpen
  \bibfield  {author} {\bibinfo {author} {\bibfnamefont {Joseph~W.}\
  \bibnamefont {Goodman}},\ }\href@noop {} {\emph {\bibinfo {title}
  {Introduction to Fourier Optics}}}\ (\bibinfo  {publisher} {W. H. Freeman},\
  \bibinfo {year} {2017})\BibitemShut {NoStop}%
\bibitem [{\citenamefont {Moustakas}(1979)}]{moustakas1979sputtered}%
  \BibitemOpen
  \bibfield  {author} {\bibinfo {author} {\bibfnamefont {TD}~\bibnamefont
  {Moustakas}},\ }\bibfield  {title} {\enquote {\bibinfo {title} {Sputtered
  hydrogenated amorphous silicon},}\ }\href@noop {} {\bibfield  {journal}
  {\bibinfo  {journal} {Journal of Electronic Materials}\ }\textbf {\bibinfo
  {volume} {8}},\ \bibinfo {pages} {391--435} (\bibinfo {year}
  {1979})}\BibitemShut {NoStop}%
\bibitem [{\citenamefont {Jansen}\ \emph {et~al.}(1997)\citenamefont {Jansen},
  \citenamefont {de~Boer}, \citenamefont {Wiegerink}, \citenamefont {Tas},
  \citenamefont {Smulders}, \citenamefont {Neagu},\ and\ \citenamefont
  {Elwenspoek}}]{jansen1997bsm}%
  \BibitemOpen
  \bibfield  {author} {\bibinfo {author} {\bibfnamefont {Henri}\ \bibnamefont
  {Jansen}}, \bibinfo {author} {\bibfnamefont {Meint}\ \bibnamefont {de~Boer}},
  \bibinfo {author} {\bibfnamefont {Remco}\ \bibnamefont {Wiegerink}}, \bibinfo
  {author} {\bibfnamefont {Niels}\ \bibnamefont {Tas}}, \bibinfo {author}
  {\bibfnamefont {Edwin}\ \bibnamefont {Smulders}}, \bibinfo {author}
  {\bibfnamefont {Christina}\ \bibnamefont {Neagu}}, \ and\ \bibinfo {author}
  {\bibfnamefont {Miko}\ \bibnamefont {Elwenspoek}},\ }\bibfield  {title}
  {\enquote {\bibinfo {title} {Bsm 7: Rie lag in high aspect ratio trench
  etching of silicon},}\ }\href@noop {} {\bibfield  {journal} {\bibinfo
  {journal} {Microelectronic Engineering}\ }\textbf {\bibinfo {volume} {35}},\
  \bibinfo {pages} {45--50} (\bibinfo {year} {1997})}\BibitemShut {NoStop}%
\bibitem [{\citenamefont {Lalanne}\ \emph {et~al.}(1999)\citenamefont
  {Lalanne}, \citenamefont {Astilean}, \citenamefont {Chavel}, \citenamefont
  {Cambril},\ and\ \citenamefont {Launois}}]{lalanne1999design}%
  \BibitemOpen
  \bibfield  {author} {\bibinfo {author} {\bibfnamefont {Philippe}\
  \bibnamefont {Lalanne}}, \bibinfo {author} {\bibfnamefont {Simion}\
  \bibnamefont {Astilean}}, \bibinfo {author} {\bibfnamefont {Pierre}\
  \bibnamefont {Chavel}}, \bibinfo {author} {\bibfnamefont {Edmond}\
  \bibnamefont {Cambril}}, \ and\ \bibinfo {author} {\bibfnamefont {Huguette}\
  \bibnamefont {Launois}},\ }\bibfield  {title} {\enquote {\bibinfo {title}
  {Design and fabrication of blazed binary diffractive elements with sampling
  periods smaller than the structural cutoff},}\ }\href@noop {} {\bibfield
  {journal} {\bibinfo  {journal} {JOSA A}\ }\textbf {\bibinfo {volume} {16}},\
  \bibinfo {pages} {1143--1156} (\bibinfo {year} {1999})}\BibitemShut {NoStop}%
\bibitem [{\citenamefont {Golub}\ and\ \citenamefont
  {Friesem}(2007)}]{golub2007analytic}%
  \BibitemOpen
  \bibfield  {author} {\bibinfo {author} {\bibfnamefont {Michael~A}\
  \bibnamefont {Golub}}\ and\ \bibinfo {author} {\bibfnamefont {Asher~A}\
  \bibnamefont {Friesem}},\ }\bibfield  {title} {\enquote {\bibinfo {title}
  {Analytic design and solutions for resonance domain diffractive optical
  elements},}\ }\href@noop {} {\bibfield  {journal} {\bibinfo  {journal} {JOSA
  A}\ }\textbf {\bibinfo {volume} {24}},\ \bibinfo {pages} {687--695} (\bibinfo
  {year} {2007})}\BibitemShut {NoStop}%
\bibitem [{\citenamefont {Rivolta}(1986)}]{rivolta1986airy}%
  \BibitemOpen
  \bibfield  {author} {\bibinfo {author} {\bibfnamefont {Claudio}\ \bibnamefont
  {Rivolta}},\ }\bibfield  {title} {\enquote {\bibinfo {title} {Airy disk
  diffraction pattern: comparison of some values of f/no. and obscuration
  ratio},}\ }\href@noop {} {\bibfield  {journal} {\bibinfo  {journal} {Applied
  optics}\ }\textbf {\bibinfo {volume} {25}},\ \bibinfo {pages} {2404--2408}
  (\bibinfo {year} {1986})}\BibitemShut {NoStop}%
\bibitem [{\citenamefont {Khorasaninejad}\ \emph
  {et~al.}(2016{\natexlab{b}})\citenamefont {Khorasaninejad}, \citenamefont
  {Chen}, \citenamefont {Zhu}, \citenamefont {Oh}, \citenamefont {Devlin},
  \citenamefont {Rousso},\ and\ \citenamefont
  {Capasso}}]{khorasaninejad2016multispectral}%
  \BibitemOpen
  \bibfield  {author} {\bibinfo {author} {\bibfnamefont {M}~\bibnamefont
  {Khorasaninejad}}, \bibinfo {author} {\bibfnamefont {WT}~\bibnamefont
  {Chen}}, \bibinfo {author} {\bibfnamefont {AY}~\bibnamefont {Zhu}}, \bibinfo
  {author} {\bibfnamefont {J}~\bibnamefont {Oh}}, \bibinfo {author}
  {\bibfnamefont {RC}~\bibnamefont {Devlin}}, \bibinfo {author} {\bibfnamefont
  {D}~\bibnamefont {Rousso}}, \ and\ \bibinfo {author} {\bibfnamefont
  {F}~\bibnamefont {Capasso}},\ }\bibfield  {title} {\enquote {\bibinfo {title}
  {Multispectral chiral imaging with a metalens},}\ }\href@noop {} {\bibfield
  {journal} {\bibinfo  {journal} {Nano letters}\ }\textbf {\bibinfo {volume}
  {16}},\ \bibinfo {pages} {4595--4600} (\bibinfo {year}
  {2016}{\natexlab{b}})}\BibitemShut {NoStop}%
\bibitem [{\citenamefont {Hoover}\ \emph {et~al.}(2011)\citenamefont {Hoover},
  \citenamefont {Young}, \citenamefont {Chandler}, \citenamefont {Luo},
  \citenamefont {Field}, \citenamefont {Sheetz}, \citenamefont {Sylvester},\
  and\ \citenamefont {Squier}}]{hoover2011remote}%
  \BibitemOpen
  \bibfield  {author} {\bibinfo {author} {\bibfnamefont {Erich~E}\ \bibnamefont
  {Hoover}}, \bibinfo {author} {\bibfnamefont {Michael~D}\ \bibnamefont
  {Young}}, \bibinfo {author} {\bibfnamefont {Eric~V}\ \bibnamefont
  {Chandler}}, \bibinfo {author} {\bibfnamefont {Anding}\ \bibnamefont {Luo}},
  \bibinfo {author} {\bibfnamefont {Jeffrey~J}\ \bibnamefont {Field}}, \bibinfo
  {author} {\bibfnamefont {Kraig~E}\ \bibnamefont {Sheetz}}, \bibinfo {author}
  {\bibfnamefont {Anne~W}\ \bibnamefont {Sylvester}}, \ and\ \bibinfo {author}
  {\bibfnamefont {Jeff~A}\ \bibnamefont {Squier}},\ }\bibfield  {title}
  {\enquote {\bibinfo {title} {Remote focusing for programmable multi-layer
  differential multiphoton microscopy},}\ }\href@noop {} {\bibfield  {journal}
  {\bibinfo  {journal} {Biomedical optics express}\ }\textbf {\bibinfo {volume}
  {2}},\ \bibinfo {pages} {113--122} (\bibinfo {year} {2011})}\BibitemShut
  {NoStop}%
\bibitem [{\citenamefont {Bewersdorf}\ \emph {et~al.}(1998)\citenamefont
  {Bewersdorf}, \citenamefont {Pick},\ and\ \citenamefont
  {Hell}}]{bewersdorf1998multifocal}%
  \BibitemOpen
  \bibfield  {author} {\bibinfo {author} {\bibfnamefont {J{\"o}rg}\
  \bibnamefont {Bewersdorf}}, \bibinfo {author} {\bibfnamefont {Rainer}\
  \bibnamefont {Pick}}, \ and\ \bibinfo {author} {\bibfnamefont {Stefan~W}\
  \bibnamefont {Hell}},\ }\bibfield  {title} {\enquote {\bibinfo {title}
  {Multifocal multiphoton microscopy},}\ }\href@noop {} {\bibfield  {journal}
  {\bibinfo  {journal} {Optics letters}\ }\textbf {\bibinfo {volume} {23}},\
  \bibinfo {pages} {655--657} (\bibinfo {year} {1998})}\BibitemShut {NoStop}%
\bibitem [{\citenamefont {Beaulieu}\ \emph {et~al.}(2020)\citenamefont
  {Beaulieu}, \citenamefont {Davison}, \citenamefont {K{\i}l{\i}{\c{c}}},
  \citenamefont {Bifano},\ and\ \citenamefont
  {Mertz}}]{beaulieu2020simultaneous}%
  \BibitemOpen
  \bibfield  {author} {\bibinfo {author} {\bibfnamefont {Devin~R}\ \bibnamefont
  {Beaulieu}}, \bibinfo {author} {\bibfnamefont {Ian~G}\ \bibnamefont
  {Davison}}, \bibinfo {author} {\bibfnamefont {K{\i}v{\i}lc{\i}m}\
  \bibnamefont {K{\i}l{\i}{\c{c}}}}, \bibinfo {author} {\bibfnamefont
  {Thomas~G}\ \bibnamefont {Bifano}}, \ and\ \bibinfo {author} {\bibfnamefont
  {Jerome}\ \bibnamefont {Mertz}},\ }\bibfield  {title} {\enquote {\bibinfo
  {title} {Simultaneous multiplane imaging with reverberation two-photon
  microscopy},}\ }\href@noop {} {\bibfield  {journal} {\bibinfo  {journal}
  {Nature Methods}\ }\textbf {\bibinfo {volume} {17}},\ \bibinfo {pages}
  {283--286} (\bibinfo {year} {2020})}\BibitemShut {NoStop}%
\bibitem [{\citenamefont {Tearney}\ \emph {et~al.}(2002)\citenamefont
  {Tearney}, \citenamefont {Bouma},\ and\ \citenamefont
  {Webb}}]{tearney2002confocal}%
  \BibitemOpen
  \bibfield  {author} {\bibinfo {author} {\bibfnamefont {Guillermo~J}\
  \bibnamefont {Tearney}}, \bibinfo {author} {\bibfnamefont {Brett~E}\
  \bibnamefont {Bouma}}, \ and\ \bibinfo {author} {\bibfnamefont {Robert~H}\
  \bibnamefont {Webb}},\ }\href@noop {} {\enquote {\bibinfo {title} {Confocal
  microscopy with multi-spectral encoding},}\ } (\bibinfo {year} {2002}),\
  \bibinfo {note} {uS Patent 6,341,036}\BibitemShut {NoStop}%
\end{thebibliography}
\end{document}